\documentclass[twoside,journey]{IEEEtran}
\usepackage{makecell}
\usepackage{hyperref}
\usepackage{array}
\usepackage{graphicx,amssymb,amsmath}
\usepackage{multicol}
\usepackage[noadjust]{cite}
\usepackage{setspace}
\usepackage{subcaption}
\usepackage{graphicx}
\usepackage{cuted} 
\usepackage{float}
\usepackage{url}
\usepackage{stfloats}
\usepackage{amsthm,pifont}
\usepackage{flushend}
\usepackage{cases,subeqnarray}
\usepackage{bm,multirow,bigstrut}
\usepackage{amsmath, amsthm, amssymb}
\usepackage{textcomp}
\usepackage{latexsym,bm}
\usepackage{booktabs}
\usepackage{xcolor}
\usepackage{mathtools}
\usepackage{dsfont}
\usepackage{extarrows}
\usepackage{epsfig}
\usepackage{epsfig}
\usepackage{epstopdf}
\usepackage{subcaption}
\usepackage[noend]{algpseudocode}
\usepackage{algorithmicx,algorithm}
\theoremstyle{plain}

\theoremstyle{plain}

\usepackage{amsmath}

\IEEEoverridecommandlockouts
\begin{document}

\title{\textit{LAMeTA}: Intent-Aware Agentic Network Optimization via a Large AI Model-Empowered Two-Stage Approach}

\author{Yinqiu Liu, Guangyuan Liu, Jiacheng Wang, Ruichen Zhang, Dusit Niyato,~\IEEEmembership{Fellow,~IEEE}, \\Geng Sun, Zehui Xiong, and Zhu Han,~\IEEEmembership{Fellow,~IEEE}
\thanks{Y. Liu G. Liu, J. Wang, R. Zhang, and D. Niyato are with the College of Computing and Data Science, Nanyang Technological University, Singapore (e-mail: yinqiu001@e.ntu.edu.sg; liug0022@e.ntu.edu.sg; jiacheng.wang@ntu.edu.sg; ruichen.zhang@ntu.edu.sg; dniyato@ntu.edu.sg). }
\thanks{G. Sun is with the College of Computer Science and Technology, Jilin University, China, and also with the College of Computing and Data Science, Nanyang Technological University, Singapore (e-mail: sungeng@jlu.edu.cn).}
\thanks{Z. Xiong is with the Pillar of Information Systems Technology and Design, Singapore University of Technology and Design, Singapore (e-mail: zehui\_xiong@sutd.edu.sg).}
\thanks{Zhu Han is with the Department of Electrical and Computer Engineering at the University of Houston, Houston, USA, and also with the Department of Computer Science and Engineering, Kyung Hee University, South Korea (e-mail: hanzhu22@gmail.com).} 
\thanks{Y. Liu and G. Liu contributed equally to the work.} 
\vspace{-0.62cm}
}
\maketitle

\begin{abstract}
Nowadays, Generative AI (GenAI) reshapes numerous domains by enabling machines to create content across modalities. 
As GenAI evolves into autonomous agents capable of reasoning, collaboration, and interaction, they are increasingly deployed on network infrastructures to serve humans automatically. 
This emerging paradigm, known as the agentic network, presents new optimization challenges due to the demand to incorporate subjective intents of human users expressed in natural language. 
Traditional generic Deep Reinforcement Learning (DRL) struggles to capture intent semantics and adjust policies dynamically, thus leading to suboptimality. In this paper, we present LAMeTA, a Large AI Model (LAM)-empowered Two-stage Approach for intent-aware agentic network optimization. First, we propose Intent-oriented Knowledge Distillation (IoKD), which efficiently distills intent-understanding capabilities from resource-intensive LAMs to lightweight edge LAMs (E-LAMs) to serve end users. Second, we develop Symbiotic Reinforcement Learning (SRL), integrating E-LAMs with a policy-based DRL framework. In SRL, E-LAMs translate natural language user intents into structured preference vectors that guide both state representation and reward design. The DRL, in turn, optimizes the generative service function chain composition and E-LAM selection based on real-time network conditions, thus optimizing the subjective Quality-of-Experience (QoE). Extensive experiments conducted in an agentic network with 81 agents demonstrate that IoKD reduces mean squared error in intent prediction by up to 22.5\%, while SRL outperforms conventional generic DRL by up to 23.5\% in maximizing intent-aware QoE.
\end{abstract}

\IEEEpeerreviewmaketitle
\vspace{-0.5cm}
\section{Introduction}
Generative AI (GenAI) has revolutionized the technological landscape, enabling machines to create content across multiple modalities, including text, images, and videos \cite{10648594}. 
Moreover, GenAI is rapidly evolving from basic content generation to complex reasoning and decision-making, transforming how machines interact with and serve humans.
This evolution has given rise to GenAI agents \cite{AgenticAI}, which represent intelligent systems that integrate generative capabilities with perception, reasoning, and autonomous problem-solving functionalities to accomplish complex tasks with minimal human intervention. 
Numerous agents collaborate in advanced communication infrastructures to deliver ubiquitous agentic services, forming agentic networks \cite{7810005, AgenticAI}. 
Recently, agentic networks have attracted great research attention \cite{AgenticAI, 10648594, 10.4018/JGIM.364094}. 
For instance, Khowaja \textit{et al.} \cite{AgenticAI} proposed a mission-critical agentic network by incorporating various agents into 6G communications. Moreover, agentic networks are also reshaping the Metaverse \cite{10.4018/JGIM.364094}, Digital Twin \cite{10648594}, and various other domains requiring intelligent, coordinated decision-making.

\begin{figure*}[tbp!]
  \centering
  \includegraphics[width=0.94\textwidth]{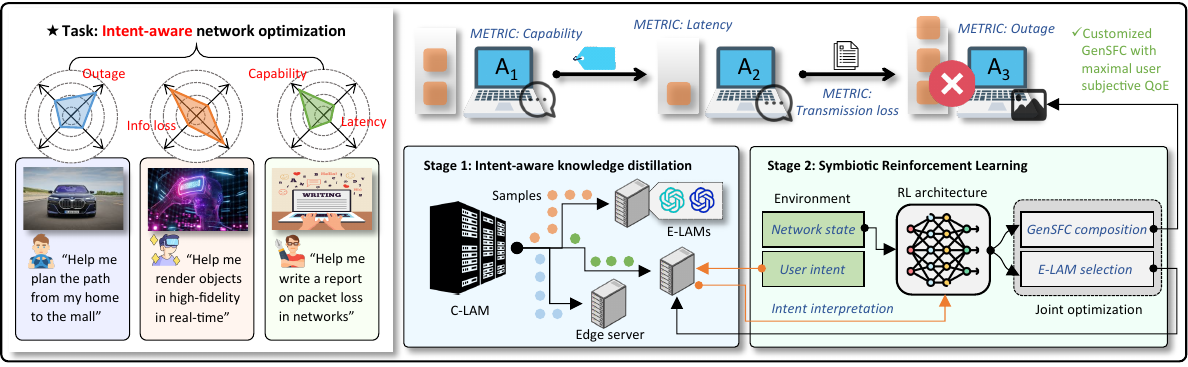} 
  \caption{The illustration of diverse applications and the corresponding preferences in the agentic network (left). The overall workflow of LAMeTA and the illustration of a $3$-step GenSFC for report writing (right). In agentic networks, the intents are incorporated into prompts explicitly or implicitly. For example, example 2 explicitly requires high capability and low latency for 3D rendering. However, the requirement for high capability (to write insightful reports) is implicitly expressed in example 3. Both explicit and implicit intents should be captured and considered.} 
  \vspace{-0.4cm}
  \label{Figure1}
\end{figure*}

To accomplish advanced tasks, service provisioning in agentic networks often involves multiple heterogeneous agents that form a Generative Service Function Chain (GenSFC) \cite{7810005, SFC1}.
Consequently, dynamically selecting the optimal agents from various candidates to maximize the efficiency of composed GenSFCs becomes a critical concern. 
Although traditional optimization methods, such as linear programming \cite{10265246} and Deep Reinforcement Learning (DRL) \cite{9869667, SFC1}, can be applied, they face a daunting challenge in agentic networks. 
As shown in Fig. \ref{Figure1}, agents are inherently developed to serve humans \cite{AgenticAI}, incorporating diverse context-driven and subjective intent factors that can hardly be represented.
Hence, relying solely on a generic DRL model or static optimization framework is inefficient in satisfying the requirements of diverse users.

Fortunately, Large AI Models (LAMs) have demonstrated remarkable potential in networking optimization \cite{10679152, 10670195}. 
Trained on massive general datasets, LAMs excel at understanding user intent through multimodal comprehension and cross-modality alignment \cite{10558819}. 
Moreover, LAMs exhibit strong zero-shot generation capabilities (i.e., directly generate optimization solutions based on user description without retraining), enabling them to tackle unseen tasks with promising performance. 
For instance, Zhang \textit{et al.} \cite{10679152} leveraged a Large Language Model (LLM) to optimize communication systems directly from natural language user descriptions.
Shao \textit{et al.} \cite{10582827} demonstrated the efficiency of LAMs in spectrum sensing and power allocation optimization.
Nonetheless, we notice that LAMs also suffer from two inherent limitations.
\begin{itemize}
    \item \textbf{Strict Resource Requirements}: LAMs significantly exceed conventional AI models in size \cite{AgenticAI}, imposing substantial computing requirements. For instance, GPT-3.5 comprises 175 billion parameters, while current mobile chips, such as \textit{Qualcomm Snapdragon}, can only support LAMs with 10 billion parameters. This contradiction limits the LAM capacity available to end users.
    \vspace{0.03cm}
    \item \textbf{Insensitivity to Numerical Conditions}: Most LAMs are optimized for understanding multimedia content, with embeddings and training oriented toward semantic representation \cite{Embedding}. However, solving optimization problems requires precise sensitivity to numerical inputs, as slight changes in environmental factors should be accurately captured by the policy to enable precise decision adjustments. Consequently, although LAMs achieve stratifying zero-shot efficiency, they cannot keep reinforcing the policies based on subtle network conditions.
\end{itemize}

To address the above challenges, in this paper, we present LAMeTA, an $\underline{\text{LAM}}$-$\underline{\text{e}}$mpowered $\underline{\text{T}}$wo-stage $\underline{\text{A}}$pproach to perform intent-aware agentic network optimization. 
First, Intent-oriented Knowledge Distillation (IoKD) addresses resource consumption issues by distilling intent-understanding capability from resource-intensive LAMs to lightweight edge LAMs (E-LAMs). 
Subsequently, Symbiotic Reinforcement Learning (SRL) overcomes the numerical insensitivity of LAMs in optimization solving by incorporating a DRL structure. 
Specifically, we consider agentic network optimization as joint GenSFC composition and E-LAM selection.
A policy network continuously interacts with the environment, infers meaningful relationships between changing conditions and optimal solutions, and progressively refines its policy. 
The selected E-LAM provides intent translation to enhance the state representation and adjust reward settings dynamically.
Consequently, the policy network can be reinforced by acquiring reward signals and optimizing the policy gradient.
Furthermore, the efficiency of the generated GenSFCs is fed back to the E-LAM, thus achieving continuous self-refinement. 
In this way, LAMeTA can efficiently provision services in varying agentic networks and maximize users' subjective experience.
Notably, our proposal is adaptable to various wireless network optimization tasks that involve complicated human participation, subjective intent factors, and context-specific requirements. Our main contributions can be summarized as follows (see Fig. \ref{Figure1}).
\begin{itemize}
    \item \textbf{Intent-aware Agentic Network Optimization}: We pre-sent LAMeTA, an LAM-empowered two-stage approach to optimize agentic networks with intents. Particularly, we model subjective Quality-of-Experience (QoE) in agentic networks and formulate a joint optimization problem of GenSFC composition and E-LAM selection. 
    \item \textbf{Intent-oriented Knowledge Distillation}: For stage 1, we propose IoKD, enabling lightweight E-LAMs to accurately predict users' subjective preferences based on their intent description. Particularly, to enhance distillation efficiency, IoKD adaptively augments samples that reflect application-specific preference patterns. Moreover, we present a weighted pairwise distillation process, which utilizes contrastive samples and weight adjustment to explicitly guide E-LAMs in distinguishing intent differences and mitigating the impact of outliers.
    \item \textbf{Symbiotic Reinforcement Learning}: For stage 2, we develop SRL, which incorporates an environmental exploration module to interact with the agentic network and policy-based RL to optimize solutions. Moreover, we integrate E-LAMs to translate intents, thus enhancing state representation and adjusting reward functions. The GenSFC performance is then fed back to the E-LAM, realizing symbiotic enhancement.
    \item \textbf{Numerical Results}: We simulate a large agentic network with 81 heterogeneous agents. Extensive experimental results demonstrate that IoKD significantly improves intent understanding precision, improving precision by up to 22.5\%. Moreover, SRL consistently outperforms generic DRL methods by 17.2-23.5\% in optimizing QoE while adapting to diverse user intents.
\end{itemize}

The remainder of this paper is structured as follows. First, Section \ref{SectionII} reviews the related work and states our motivations. The system models and problem formulation are described in Section \ref{SectionIII}. Then, Sections \ref{SectionIV} and \ref{SectionV} present IoKD and SRL, respectively. The numerical results are discussed in Section \ref{SectionVI}. Finally, Section \ref{SectionVII} concludes this paper.

\section{Related Works and Motivation}\label{SectionII}
\subsection{Agentic Networking}
Agentic network research has attracted great attention. 
For instance, Liu \emph{et al.} \cite{Opt3} presented DyLAN, which introduces different topologies to organize multiple agents and dynamically forms agent teams based on specific user queries to maximize collaboration efficiency.
Oriented to 6G networks, Xiao \emph{et al.} \cite{AgenticNet1} presented AgentNet, a framework that facilitates efficient inter-agent communications and knowledge transfers. 
Similarly, Dev \emph{et al.} \cite{AgentNet2} explored next-generation wireless communications empowered by agents, discussing how to deploy agents in resource-limited and serverless computing networks. 
Xu \emph{et al.} \cite{10648594} introduced a split learning scheme for LLM agents in 6G networks, partitioning agent functions between mobile and edge devices to improve network-based perception and resource efficiency. 
Khowaja \emph{et al.} \cite{AgenticAI} presented a multi-layer agentic AI framework for mission-critical 6G applications, focusing on decentralized decision-making and adaptive resource allocation.
Finally, Abdelnabi \emph{et al.} \cite{AgenticNet2} proposed a firewall-based framework that dynamically secures multi-turn LLM agent communications.
Despite such progress, service provisioning in agentic networks remains under-researched. 
Unlike traditional network scenarios, agents are designed to assist users autonomously, necessitating careful consideration of user intent as a critical factor.



\subsection{LAM-based Intent Understanding}

Trained on vast corpora, LAMs excel in linguistic understanding and generation, allowing users to express complicated intent by natural language from an infinite vocabulary.
The authors in \cite{10574890, 10539172, 10575237} leveraged LAMs to translate user intents into specific network routing, deployment, and function virtualization, respectively.
Kou \emph{et al.} \cite{10902595} proposed an adaptive LAM, which first comprehends the user intents and iteratively refines its translation capability from the simulation results.
Lira \emph{et al.} \cite{10726902} fine‑tuned LAM to autonomously convert maintenance intents into comprehensive zero‑touch configuration updates about router forwarding tables, dynamic interface reconfigurations, and QoE parameter adjustments. 
In terms of security, Dzeparoska \emph{et al.} \cite{10327837} utilized LLMs to interpret application‑layer intents involving the restriction of access to sensitive services and to generate authentication policies accordingly. 
Finally, LAMs facilitate network optimization by dynamically allocating resources based on user intents.
Habib \emph{et al.} \cite{Opt2} presented an LAM-enhanced resource allocation scheme, in which LAMs are leveraged to classify user intent and feed extracted terms into the DRL.
Thus, resource allocations can be optimized by tailoring them to specific user demands.
Inspired by such progress, LAMeTA presents IoKD to enhance intent understanding of LAMs and achieves intent-aware optimization.

\subsection{LAM for Network Optimization}
LAMs can serve as a powerful zero-shot/few-shot network optimizer \cite{10829820,10685369}. 
First, optimization problems can be directly expressed through natural language and solved by LAMs given their powerful context understanding capabilities. 
For instance, the authors in \cite{10681550, Opt4, Opt1} expressed wireless access point placement, resource allocation, and power control problems by prompts, which are fed to LAMs to generate solutions.
Beyond direct optimization generation, LAMs can be refined through domain-specific knowledge adaptation or fine-tuning to enhance their capabilities further. 
For example, Kan \emph{et al.} \cite{10583947} presented MobileLLaMA, an instruction-finetuned LLM that generates codes to optimize packet inspection and IP routing. 
Shao \emph{et al.} \cite{10582827} presented WirelessLLM, a domain‐adapted framework that fuses wireless-specific knowledge within LLMs to optimize power allocation, spectrum sensing, etc. 
Huang \emph{et al.} \cite{10.1109/MNET.2024.3435752} proposed ChatNet, which adopts four LAMs to perform end-to-end requirement analysis, optimization planning, QoE calculation, and policy execution. 
Finally, LAMs can process multimodal information that affects optimization objectives, enabling more comprehensive network solutions beyond traditional data types.
Jiang \emph{et al.} \cite{10670195} utilized LLMs to extract, align, and recover multimodal semantic information of semantic communications, thus optimizing communication efficiency.

Nonetheless, LAMs cannot efficiently observe the numerical network states and keep reinforcing their policy, which often leads to suboptimality. 
To this end, we present SRL, which synergistically combines the semantic understanding capabilities of LAMs with the numerical optimization strengths of DRL architecture.



\section{System Model: Agentic Network with Intent-aware Optimization}\label{SectionIII}
In this section, we first model the agentic network environment and QoE factors.
Afterward, we formulate the problem of intent-aware genetic network optimization.
Finally, we present the architectural design of our LAMeTA.

\subsection{Agentic Network and Services}
We consider an agentic network with $N$ GenAI agents $\{A_1, A_2, \dots, A_N\}$, each operating a GenAI model (usually an LAM) to perform a specific type of function. 
These agents establish connections based on collaborative requirements and underlying network connectivity.
Hence, we use an undirected graph $\mathcal{G} = (\mathcal{V}, \mathcal{E})$ to model such a network topology, where the vertices $\mathcal{V}$ and edges $\mathcal{E}$ denote the sets of agents and communication links, respectively.

Users submit service requests to the agentic network. 
Fig. \ref{Figure1} illustrates an example service request and the corresponding GenSFC. 
In detail, a user requests the agentic network to write a technical report about packet loss in networks. 
Accordingly, the GenSFC should comprise three sequential agents: an agent with network access and reasoning capabilities to gather and analyze related documents, an LLM for report writing, and a multimodal LAM with statistical modules to sort data and draw figures and tables. 
Note that the user prompt implicitly requires maximizing the generation quality. 
The objective of LAMeTA is to optimize agentic networks to serve users with various requests and intents.

\subsection{Objective QoE Modeling}
In this part, we model the QoE of users with the generated GenSFC. Specifically, we involve four factors that are highly oriented to agentic services, namely, agent capability, information loss, service latency, and service outage probability.
\subsubsection{Agent Capability}
First, we consider the capability of each agent $A_i,\, i \in \{1, 2, \dots, N\}$ in GenSFC to perform assigned tasks. 
Unlike traditional computing architectures \cite{10032267} where the capability can be reflected directly by the computational power assigned, agentic services are based on LAMs, whose resource investment and resulting performance exhibit a complicated pattern. 
Inspired by LAM scaling laws \cite{10.5555/3600270.3602446}, we model the agent capability (determined by the performance of its LAM acquired in the training stage) considering computation budget $C_i$, available training tokens $|\mathcal{D}_i|$, and the number of parameters $|\mathcal{N}_i|$, 
Specifically, given $C_i$, the resulting pre-training loss can be expressed as \cite{10.5555/3600270.3602446}
\begin{subequations}
\begin{flalign}
\mathcal{L}^i_{\text{pre-training}}(|\mathcal{N}_i|, |\mathcal{D}_i|) &= \mathcal{L}_0 + \frac{\zeta}{|\mathcal{N}_i|_\text{OP}^{\alpha_1}} + \frac{\eta}{|\mathcal{D}_i|_\text{OP}^{\alpha_2}},\\
|\mathcal{N}_i|_{\text{OP}} &= \left(\frac{\alpha_1\zeta}{\alpha_2\eta}\right)^{\frac{1}{\alpha_1+\alpha_2}}\left(\frac{C_i}{6}\right)^{\frac{\alpha_2}{\alpha_1 + \alpha_2}}, \\
|\mathcal{D}_i|_{\text{OP}} &= \left(\frac{\alpha_2\eta}{\alpha_1\zeta}\right)^{\frac{1}{\alpha_1+\alpha_2}}\left(\frac{C_i}{6}\right)^{\frac{\alpha_1}{\alpha_1 + \alpha_2}}, 
\end{flalign}
\end{subequations}
where $|\mathcal{N}_i|_\text{OP}$ and $|\mathcal{D}_i|_\text{OP}$ represent the optimal numbers of LAM parameters and training tokens, respectively. $\mathcal{L}^i_{\text{pre-training}}$ consists of three components: i) $\mathcal{L}_0$ captures the loss of an ideal generative process, which can be regarded as 0, ii) $\frac{\zeta}{|\mathcal{N}_i|_\text{OP}^{\alpha_1}}$ captures the fact that an LAM with $|\mathcal{N}_i|_{\text{OP}}$ parameters underperforms the ideal generative process (with infinite parameters), and iii) $\frac{\eta}{|\mathcal{D}_i|_\text{OP}^{\alpha_2}}$ captures the fact that the LAM is not trained to convergence, since we only perform a finite number of optimization steps using $|\mathcal{D}_i|_\text{OP}$ samples. According to \cite{10.5555/3600270.3602446}, the values of $\zeta$, $\eta$, $\alpha_1$, and $\alpha_2$ are empirically set as {406.4, 410.7, 0.34, and 0.28}, respectively.
With pre-training loss $\mathcal{L}^i_{\text{pre-training}}$, the probability that $A_i$ successfully completes the task can be modeled as $e^{(-\mathcal{L}^i_\text{pre-training})}$ \cite{10.5555/3600270.3602446}. 
Consequently, the capability of the entire $n$-step GenSFC is defined as
\begin{equation}
    C_\text{GenSFC} = \frac{1}{n}\prod_{i=1}^{n}\text{exp}(-\mathcal{L}^{i}_{\text{pre-training}}).
\end{equation}
Note that agent capability is mainly determined by the LAM training process. In contrast, the computational power assigned for inference primarily affects the processing latency.

\subsubsection{Information Loss}
Information loss between agents in a GenSFC is particularly critical, since the prompt for the next agent is the output of the previous agent (see Fig. \ref{Figure1}). 
LAMs are inherently sensitive to prompts since they serve as both instructions and context for generation \cite{10670195}. 
Slight errors in prompt transmission can cause significant semantic differences or quality drops. 
Hence, we derive the total information loss by Bit Error Rate (BER) \cite{5654629} between adjacent agents in the GenSFC.
Suppose that the BER of the communication links connecting agents $A_i$ and $A_j$ is $\mathbb{E}[\textit{BER}_j]$, the BER of the entire $n$-step GenSFC can be expressed as
\begin{equation}
    \textit{B}_\text{GenSFC} = 1 - P_\text{correct} = 1 - \prod_{i=2}^{n} \left(1-\mathbb{E}[\textit{BER}_i]\right).
\end{equation}

\subsubsection{Service Latency}
Service latency is another critical factor affecting user QoE in agentic networks, as excessive latency significantly degrades user experience. 
We model each GenSFC as an M/G/1 queuing system \cite{WANG20041713} and formulate the service latency accordingly.
Specifically, we suppose that service requests arrive according to a Poisson process with rate $\lambda$. 
The service time of $A_i$ follows a general distribution with mean $\mathbb{E}[S_i] = \frac{1}{\mu_i}$ and variance $\sigma_i^2$. 
Each agent provides services in a first-come, first-served manner with a single thread.
Finally, the completion of service at agent $A_i$ triggers the arrival of the next agent in the GenSFC.

For a given $n$-step GenSFC, we analyze end-to-end latency by examining the contribution of each agent to the overall service time. First, the traffic intensity at agent $A_i$ is:
\begin{equation}
\rho_i = \frac{\lambda}{\mu_i}, \rho_i < 1, \quad \forall i \in \{1, 2, \ldots, n\}.
\end{equation}
The traffic intensity represents the ratio between the arrival rate and the service rate, indicating the average utilization of the agent.
For system stability across the agentic network, we require $\rho_i < 1$.
This constraint ensures that each agent can process requests faster than they arrive, preventing infinite queue buildup.
The first moment of service time at agent $A_i$ is $\mathbb{E}[S_i] = \frac{1}{\mu_i}$. The second moment, which captures both the mean and variance of service time, is $\mathbb{E}[S_i^2] = \sigma_i^2 + \left(\frac{1}{\mu_i}\right)^2$.
The coefficient of variation, representing the normalized service time variability of agent $A_i$, is then defined as the ratio of standard deviation to mean:
\begin{equation}
V_i = \sigma_i\mu_i = \frac{\sigma_i}{\mathbb{E}[S_i]}.
\end{equation}
Using the Pollaczek-Khinchine formula \cite{WANG20041713}, the mean waiting time in queue at agent $A_i$ is:
\begin{equation}
W^q_i = \frac{\lambda\mathbb{E}[S_i^2]}{2(1-\rho_i)} = \frac{\rho_i(1+V_i^2)}{2\mu_i(1-\rho_i)}.
\end{equation}
Therefore, the mean latency (including both the service and waiting time) at agent $A_i$ is:
\begin{equation}
L_i = \frac{1}{\mu_i} + W^q_i = \frac{1}{\mu_i}\left(1 + \frac{\rho_i(1+V_i^2)}{2(1-\rho_i)}\right).
\end{equation}

In conclusion, the end-to-end service latency across the entire $n$-step agentic GenSFC is\footnote{For agentic AI services, service and waiting time dominate the overall service latency due to the computation-intensive nature of generative tasks. Therefore, we focus on these components and consider the transmission latency to be negligible.}:
\begin{equation}
L_{\text{GenSFC}} = \sum_{i=1}^{n} L_i = \sum_{i=1}^{n} \frac{1}{\mu_i}\left(1 + \frac{\rho_i(1+V_i^2)}{2(1-\rho_i)}\right).
\end{equation}

\subsubsection{Service Outage Probability}
Another issue that affects QoE is the possible service outage of a GenSFC. We consider two primary failure scenarios that may lead to service disruption: i) the crash of an individual agent within the GenSFC, and ii) service denial when the incoming workload exceeds the GenSFC's processing capacity.

First, the probability that the GenSFC operates normally without any agent outage can be expressed as:
\begin{equation}
P_{\text{no-crash}} = \prod_{i=1}^{n} \left(1-\frac{|F_i|}{|O_i|}\right),
\end{equation}
where $|O_i|$ and $|F_i|$ represent the total number of observations and the number of events that $A_i$ is in outage (caused by hardware failures, software errors, etc.), respectively.

To model the probability of service denial due to excessive workload, we first analyze the maximum sustainable throughput of the GenSFC. Under heavy traffic conditions when $\rho_i \rightarrow$ 1 (i.e., $\lambda$ $\rightarrow$ $\mu_i$), we can observe that waiting time increases dramatically, as
\begin{equation}
    L_i = \frac{1}{\mu_i} + \frac{\lambda\mathbb{E}[S_i^2]}{2\left(\frac{\mu_i-\lambda}{\mu_i}\right)} = \frac{1}{\mu_i} + \frac{\lambda\mathbb{E}[S_i^2]\mu_i}{2(\mu_i-\lambda)}.
    \label{Eq10}
\end{equation}
Note that the derivation in (\ref{Eq10}) is achieved by substituting $1-\rho_i = 1-\frac{\lambda}{\mu_i} = \frac{\mu_i-\lambda}{\mu_i}$.
Clearly, if $\lambda \rightarrow \min\{u_i\}, \forall i \in \{1, 2, \dots, n\}$, $W_i \rightarrow \infty$. This causes the GenSFC service outage.
Hence, to identify the critical bottleneck in the GenSFC, we locate the agent with the highest traffic intensity:
\begin{equation}
i^{*} = \arg\max_{i \in \{1,2,\ldots,n\}} \{\rho_i\}.
\end{equation}
The agent $A_{i^{*}}$ represents the bottleneck in GenSFC and becomes increasingly important as traffic intensifies.
Accordingly, the maximum sustainable throughput of the entire GenSFC is constrained by the agent with the lowest service rate:
\begin{equation}
\lambda_{\max} = \mu_{i^{*}}.
\label{Eq12}
\end{equation}
(\ref{Eq12}) defines the fundamental throughput limit of the GenSFC. Attempting to process requests at a rate exceeding $\lambda_{\max}$ will inevitably lead to GenSFC buildup and service outage.
Recall that user requests follow a Poisson process with rate $\lambda$, the probability of observing more than $\lambda_{\max}$ requests in a unit time interval can be derived as:
\begin{equation}
P_{\text{over}} = P(R > \lambda_{\max}) = 1 - \sum_{k=0}^{\lfloor\lambda_{\max}\rfloor} \frac{e^{-\lambda}\lambda^k}{k!},
\end{equation}
where $R$ represents the random variable for the number of arrivals per unit time.
Combining both failure scenarios, the probability that the GenSFC operates normally is:
\begin{equation}
P_{\text{nor}} = P_{\text{no-crash}} \cdot (1 - P_{\text{over}}) = \prod_{i=1}^{n} \left(1-\frac{|F_i|}{|O_i|}\right)\! \left(\sum_{k=0}^{\lfloor\lambda_{\max}\rfloor} \frac{e^{-\lambda}\lambda^k}{k!}\right)\!.
\end{equation}
Consequently, the overall probability of service outage $P_\text{outage}$ = $1 - P_\text{nor}$.
This comprehensive model captures both the reliability aspects of individual agents and the capacity constraints of the overall system, providing a realistic assessment of service availability in agentic networks.

\subsubsection{QoE Modeling}
Combining four QoE factors, the overall user QoE toward a given GenSFC can be defined as
\begin{equation}
\begin{split}
    \mathcal{Q}(&C_\text{GenSFC}, B_\text{GenSFC}, W_\text{GenSFC}, P_\text{outage}) \!=\!\\ (1-P_\text{outage})&\left((1-B_\text{GenSFC})\,\text{ln}\left(\frac{C_\text{GenSFC}}{C_\text{th}}\right) - \text{ln}\left(L_\text{GenSFC}\right)\right),
    \end{split}
    \label{Eq15}
\end{equation}
where $C_\text{th}$ denotes the user capability threshold.
Particularly, to effectively model the user's QoE toward service latency and generation quality, we apply the Weber-Fechner law \cite{10144339}. 
This law states that as the stimulus to users (e.g., agent capability) increases, the perceived sensation grows but at a diminishing rate, which can be modeled as a logarithmic relationship. 
\vspace{-0.5cm}
\subsection{Subjective User Intent and Problem Formulation}
Traditionally, the QoE function, such as (\ref{Eq15}), can be adopted directly as an optimization objective. 
However, in human-oriented scenarios represented by agentic networks, applying a unified QoE measure is inadequate.
This inadequacy stems from the fact that user-specific subjective preferences and task-related intents are overlooked.

We consider the case in Fig. \ref{Figure1} as an example.
The user task description and prompt implicitly indicate that $C_\text{GenSFC}$ and $B_\text{GenSFC}$ are prioritized over other factors. 
This preference arises because report writing is particularly sensitive to GenAI capability (to maximize generation quality) and input accuracy (to guarantee that user semantics and generated materials are transmitted along the GenSFC with the minimal error). 
In contrast, for real-time applications, such as 3D environment rendering, service latency becomes the dominant concern. 
Affected by such task-specific subjective intent, user-perceived QoE differs from the unified QoE defined in (\ref{Eq15}), which can be expressed as
\begin{equation}
\begin{split}
    \mathcal{Q}(C_\text{GenSFC}, B_\text{GenSFC}, W_\text{GenSFC}, P_\text{outage}) &\xrightarrow{f_\theta(\mathcal{V}^{\infty}) = \mathbf{s}}\\\mathcal{Q}^{*}(C_\text{GenSFC}, B_\text{GenSFC}, W_\text{GenSFC}, & P_\text{outage}, \mathbf{s}),
\end{split}
\end{equation}
where $f_\theta(\cdot)$ denotes an implicit function that maps the objective QoE to user perception based on the specific intent, and $\mathcal{V}^{\infty}$ represents the infinite vocabulary space from which user intents are expressed. The vector $\mathbf{s}$ captures subjective factors generated by $f_\theta(\mathcal{V}^{\infty})$.
Formulating this transformation faces significant challenges due to two-fold reasons. 
First, the function $f_\theta(\cdot)$ reflects complex cognitive and perceptual processes, making it difficult to formulate or approximate. Second, the input intent encompasses natural language expressions drawn from an infinite vocabulary, creating a vast input space.

To enable intent-aware agentic network optimization, we introduce LAMs to learn and approximate the complex mapping function $f_\theta(\mathcal{V}^{\infty})$. 
As aforementioned, LAMs are particularly well-suited for intent understanding since their extensive pre-training on numerous text corpora enables them to interpret natural language from $\mathcal{V}^{\infty}\!$. 
Additionally, LAMs excel at capturing implicit relationships between linguistic expressions and their underlying semantic meanings, making them ideal for modeling the complicated transformation between user intent and quality perception. 
Without loss of generality, we define the subjective factor $\mathbf{s}$ as:
\begin{equation}
    \mathbf{s} \triangleq [\omega_C, \omega_B, \omega_L, \omega_P],
\end{equation}
where $\omega_C$, $\omega_B$, $\omega_L$, and $\omega_P$ are weighting factors corresponding to capability, information loss, service latency, and service outage probability, respectively. 
We can observe that user intent affects the weight distribution among different QoE factors, reflecting how users inherently value each aspect of QoE based on their specific applications and preferences. 
In this case, we further define the intent-aware agentic network optimization problem as
\begin{subequations}
    \begin{flalign}
            \max_{\{\mathcal{G}_\text{GenSFC}, \,\kappa\}} \;&\mathcal{Q}^{*}\,(C_\text{GenSFC}, B_\text{GenSFC}, L_\text{GenSFC}, P_\text{outage}, \mathbf{s}_\kappa) - \mathcal{F}(\kappa),&& \label{eq:opt_a} \\
            &s.t. ~~~~ L_\text{GenSFC} \leq L_\text{max}, && \label{eq:opt_b}\\
            &~~~~~~~~C_\text{GenSFC} \geq C_\mathrm{th}, && \label{eq:opt_c}\\
            &~~~~~~~~\mathrm{Succ(}\mathcal{G}_\text{GenSFC}) =  \text{True}, && \label{eq:opt_d}
    \end{flalign}
\end{subequations}
where $\mathcal{G}_\text{GenSFC}$ = ($\mathcal{V}_\text{GenSFC}$, $\mathcal{E}_\text{GenSFC}$), denoting the generated GenSFC.
$\kappa$ denotes the LAM selected for understanding intent, with $\mathbf{s}_\kappa$ representing the translated subjective factor.
$\mathcal{F}(\kappa)$ represents the service fee of calling LAM $\kappa$, which is correlated to the energy consumption of $\kappa$.
(\ref{eq:opt_b}) requires that the service latency cannot exceed the maximum tolerated time. 
(\ref{eq:opt_c}) means that the capability of GenSFC should at least reach the user threshold. 
Otherwise, the service execution will be considered a failure.
(\ref{eq:opt_d}) requires that the generated GenSFC should match the user's functional requirements. 
From (\ref{eq:opt_a}), we can observe that we consider a joint optimization problem on both GenSFC composition and LAM selection.
Next, we present LAMeTA to solve this problem.
\begin{figure}[tbp!]
  \centering
  \includegraphics[width=0.48\textwidth]{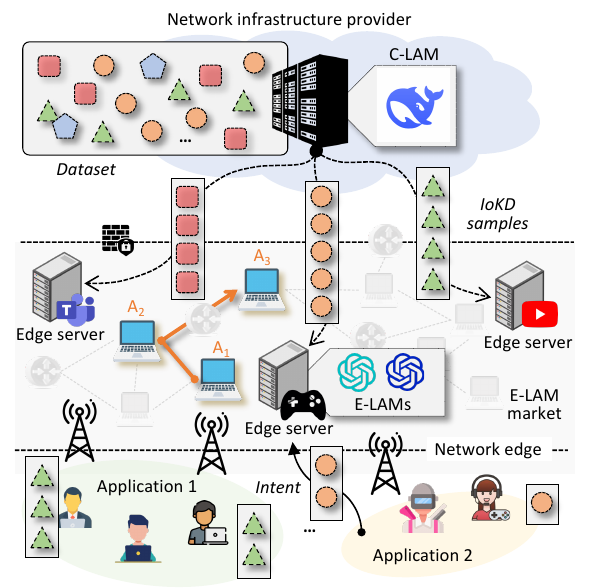} 
  \vspace{-0.1cm}
  \caption{The illustration of LAMeTA for agentic network optimization. The items with different shapes and colors represent intents belonging to different applications. We can observe that applications 1 and 2 refer to business and gaming, respectively.} 
  \vspace{-0.2cm}
  \label{Figure2}
\end{figure}

\subsection{LAMeTA Design Overview}
Fig. \ref{Figure2} illustrates the architecture of LAMeTA. A centralized LAM (C-LAM), such as Deepseek R1 or GPT-4 with hundreds of billions of parameters, is deployed by the network infrastructure provider. 
Serving as a central coordinator, the C-LAM monitors the entire agentic network and exhibits superior comprehension and generation capabilities.
However, the influx of users requesting services from the resource-intensive C-LAM often leads to service congestion, increased latency, and potential privacy risks due to the single point of failure. 
To this end, multiple edge servers are deployed, each of which serves nearby users within a specific application domain.
As depicted in Fig. \ref{Figure2}, a market of E-LAM services is established.
First, users can flexibly choose suitable edge servers for GenSFC composition.
Furthermore, each edge server hosts multiple E-LAMs with varying model sizes, intent understanding capabilities, and associated costs to serve users with different budgets. 
In the following section, we introduce IoKD, which distills intent understanding capabilities from the C-LAM to E-LAMs. 
Subsequently, SRL addresses the joint optimization of GenSFC composition and E-LAM selection, enabling edge servers to generate customized GenSFCs.
\begin{figure*}[tbp!]
  \centering
  \includegraphics[width=0.93\textwidth]{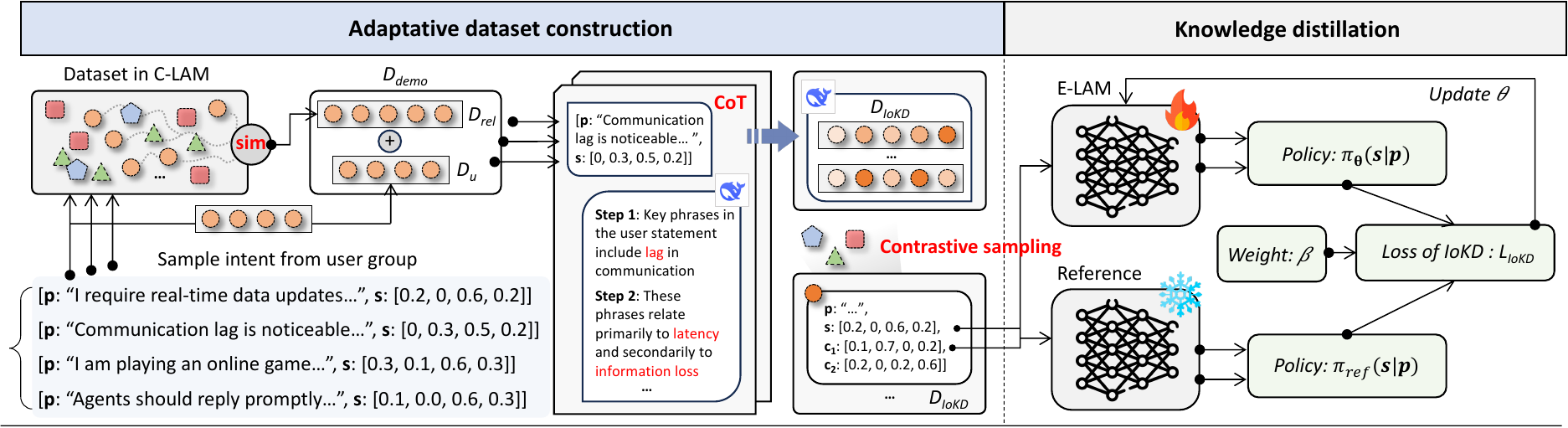} 
  \vspace{-0.12cm}
  \caption{The illustration of IoKD. The left part demonstrates the process of adaptive dataset construction, which is the prerequisite of IoKD. The right part demonstrates the IoKD process.} 
  \vspace{-0.5cm}
  \label{Figure3}
\end{figure*}

\section{Intent-oriented Knowledge Distillation}\label{SectionIV}

\subsection{Adaptive Dataset Construction}
To teach lightweight E-LAMs to understand user intents within an application domain, sufficient application-specific $\langle$\textit{prompt}, \textit{subjective factor}$\rangle$ pairs are required. 
To this end, IoKD leverages C-LAM to construct such datasets adaptively. 
First, the C-LAM monitors network information and maintains historical records, which can be expressed as:
\begin{equation}
    \mathcal{D}_\text{h} = \left[d^1_\text{h} = \langle \mathbf{p}^1_\text{h}, \mathbf{s}^1_\text{h}\rangle, \,d^2_\text{h}, \,\dots, \,d^{|\mathcal{D}_h|}_\text{h}\right],
\end{equation}
where $\mathbf{p}_\text{h}$ denotes the user's prompts that explicitly or implicitly describe the intents.
Since $\mathcal{D}_\text{h}$ contains diverse patterns across all applications, the number of samples that specifically match the target application is limited.
Given the strong few-shot generation capability of C-LAM, we utilize it to augment more samples.
Denoting the samples from users using the target application as $\mathcal{D}_\text{u} = \{d^1_\text{u} = \langle\mathbf{p}^1_\text{u}, \mathbf{s}^1_\text{u}\rangle, \,d^2_\text{u},\, \dots, d^{|\mathcal{D}_u|}_\text{u}\}$, the C-LAM performs the following four steps (see Fig. \ref{Figure3}).

\subsubsection{Semantic Similarity Filtering}
Based on $\mathcal{D}_\text{u}$, the C-LAM traverses $\mathcal{D}_\text{h}$ and retrieves all relevant items through semantic similarity. Particularly, we adopt the Sentence-BERT (SBERT) \cite{SBERT}, an embedding-based cosine similarity metric, which captures semantic relationships more effectively than traditional term-based approaches, i.e.,
\begin{subequations}
\begin{align}
    &\mathcal{D}_\text{rel} = \left\{d^j_\text{h} \in \mathcal{D}_\text{h} \mid \max_{i} \text{sim}(\mathbf{p}^i_\text{u}, \mathbf{p}^j_\text{h}) \geq \tau_\text{min}\right\}, &&\label{eq20a}\\
    \text{s}&\text{im}(\mathbf{p}^i_\text{u}, \mathbf{p}^j_\text{h}) = \frac{\text{SBERT}(\mathbf{p}^i_\text{u}) \cdot \text{SBERT}(\mathbf{p}^j_\text{h})}{\|\text{SBERT}(\mathbf{p}^i_\text{u})\| \cdot \|\text{SBERT}(\mathbf{p}^j_\text{h})\|}, &&\label{eq20b}
\end{align}
\end{subequations}
where $i \in \{1, 2, \dots, |\mathcal{D}_\text{u}|\}$ and $j \in \{1, 2, \dots, |\mathcal{D}_\text{h}|\}$.
(\ref{eq20b}) measures cosine similarity between the semantic embeddings of users' prompts and historical prompts, with $\tau_\text{min}$ serving as a similarity threshold that controls the quality-quantity trade-off of the retrieved samples. This approach ensures that only semantically relevant historical samples that match the user's intent patterns are selected.

\subsubsection{Chain-of-Thought Construction}
Combining $\mathcal{D}_\text{u}$ and the retrieved historical samples $\mathcal{D}_\text{rel}$, we construct a demonstration dataset for IoKD:
\begin{equation}
    \mathcal{D}_\text{demo} = \mathcal{D}_\text{u} \cup \mathcal{D}_\text{rel}.
\end{equation}
We then construct chain-of-thought reasoning paths between input prompts $\mathbf{p}$ and resulting subjective intent vectors $\mathbf{s}$ using the LAM. For each sample $d^i_\text{demo} = \langle\mathbf{p}^i_\text{demo} , \mathbf{s}^i_\text{demo}\rangle$ $\in \mathcal{D}_\text{demo}$, $i \in \{1, 2, \dots, |\mathcal{D}_\text{demo}|\}$, we can acquire
\begin{equation}
    \mathcal{R}(d^i_\text{demo} )= \text{C-LAM}\left(f_{\text{CoT}}(d^i_\text{demo})\right),
\end{equation}
where $f_{\text{CoT}}(d^i_\text{demo})$ formulates a prompt that instructs the C-LAM to explain why $\mathbf{p}^i_{\text{demo}}$ can lead to $\mathbf{s}^i_{\text{demo}}$, and $\mathcal{R}(\cdot)$ represents the chain-of-thought reasoning path. This approach increases the reliability of C-LAM for augmenting datasets by making the generation process transparent. By showcasing the complete reasoning path from prompt semantics to subjective weights to C-LAM, it is less likely to produce arbitrary or inconsistent samples during the augmentation.

\subsubsection{Dataset Augmentation}
Using $\mathcal{D}_\text{demo}$ as few-shot learning examples, we employ the C-LAM to generate additional synthetic samples to enrich the training dataset:
\begin{equation}
    \mathcal{D}_\text{IoKD} = \mathcal{D}_\text{demo} \cup \left\{d^k_\text{syn}=\langle\mathbf{p}^k_\text{syn}, \mathbf{s}^k_\text{syn}\rangle | k \in [1, N_\text{aug}]\right\}.
\end{equation}
The number of augmented samples $N_\text{aug}$ is determined based on the complexity of the intent patterns and the number of samples available in $\mathcal{D}_\text{demo}$. 

\subsubsection{Contrastive Sampling}
To effectively teach E-LAMs to predict subjective intent based on prompt semantics, not only positive but also contrastive samples should be provided. In this way, E-LAMs can be tuned to maximize the alignment with target subjective vectors while minimizing undesired associations, thus improving the distillation efficiency.

Inspired by reject sampling techniques, we establish a contrastive learning framework.
Specifically, for each $d^i_{\text{IoKD}} \in \mathcal{D}_\text{IoKD}$, we find the most divergent subjective vectors from $\mathcal{D}_\text{h}$.
Note that divergence is measured by angular distance, i.e.,
\begin{equation}
    \text{Dis}(\mathbf{s}^i_{\text{IoKD}}, \mathbf{s}^j_{\text{h}}) = \frac{\arccos\left(\frac{\mathbf{s}^i_{\text{IoKD}} \cdot \mathbf{s}^j_{\text{h}}}{\|\mathbf{s}^i_{\text{IoKD}}\| \cdot \|\mathbf{s}^j_{\text{h}}\|}\right)}{\pi},
    \label{Eq24}
\end{equation}
where $\mathbf{s}^i_{\text{IoKD}} = [\omega_C^i, \omega_B^i, \omega_W^i, \omega_P^i]$ is the subjective vector within sample $d^i_{\text{IoKD}}$, $\mathbf{s}^j_{\text{h}} = [\omega_C^j, \omega_B^j, \omega_W^j, \omega_P^j]$ is the subjective vector of the potential contrastive sample $d^j_{\text{h}} \in \mathcal{D_\text{h}}$, $\arccos(\cdot)$ is the inverse cosine function that converts the cosine similarity to an angle in radians, and $\|\cdot\|$ denotes the Euclidean. 

For each sample in $\mathcal{D}_\text{IoKD}$, we perform top-$k$ selection to identify the most effective contrastive examples:
\begin{equation}
    \mathcal{C}^i_{\text{IoKD}} = \text{Top-}k\left( \text{Dis}(\mathbf{s}^i_{\text{IoKD}}, \mathbf{s}^j_{\text{h}}), \forall \mathbf{s}^j_{\text{h}} \in \mathcal{D}_\text{h} \right),
\end{equation}
Finally, the well-prepared $\mathcal{D}_\text{IoKD}$ dataset can be formatted as
\begin{equation}
    \mathcal{D}_\mathrm{IoKD} =\! \left\{d^i_{\text{IoKD}}\! =\! \langle \mathbf{p}^i_{\text{IoKD}}, \mathbf{s}^i_{\text{IoKD}}, \mathcal{C}^i_{\text{IoKD}} \rangle | i \in \{1, \dots, |\mathcal{D}_\text{IoKD}| \} \right\}.
\end{equation}
The capability of C-LAM to understand user intents is preserved in $\mathcal{D}_\text{IoKD}$.
Note that when the number of samples within a specific application domain is sufficient, the C-LAM can adaptively reduce the degree of augmentation.
Moreover, after policies targeting major applications are well-trained, users do not need to upload intents to train new policies but can directly select the most aligned edge server.

\vspace{-0.5cm}
\subsection{Knowledge Distillation Process}
Traditional knowledge distillation approaches typically enforce the student model to directly mimic the output probability distribution of the teacher model. 
However, we intend to enable E-LAMs to precisely predict user preferences from intents rather than acquiring the full representation capability of the C-LAM. 
Inspired by Direct Preference Optimization (DPO) \cite{DPO}, IoKD presents a weighted pairwise distillation approach that focuses on subjective vector predictions.
Specifically, IoKD fetches samples from $\mathcal{D}_\text{IoKD}$, acquiring prompt $\mathbf{p}$, the corresponding subjective factor $\mathbf{s}_p$, and the contrastive factor $\mathbf{s}_c \in \mathcal{C}$. The prediction of E-LAM should align with $\mathbf{s}_p$ rather than $\mathbf{s}_c$.
Such a relationship can be expressed using the Bradley-Terry model \cite{DPO}:
\begin{subequations}
\begin{flalign}
    P(\mathbf{s}_p \succ \mathbf{s}_c | \mathbf{p}) &= \frac{\exp(r(\mathbf{p}, \mathbf{s}_p))}{\exp(r(\mathbf{p}, \mathbf{s}_p)) + \exp(r(\mathbf{p}, \mathbf{s}_c))}, \\&= \sigma(r(\mathbf{p}, \mathbf{s}_p) - r(\mathbf{p}, \mathbf{s}_c)),
\end{flalign}
\end{subequations}
where $r(\ast, \diamond)$ is a reward function quantifying the appropriateness of subjective vector $\diamond$ for prompt $\ast$, and $\sigma(x) = \frac{1}{1+\exp(-x)}$ is the logistic function.

Directly modeling $r(\ast, \diamond)$ requires extensive labeled data to approximate the relationship between prompts and subjective vectors. Hence, we establish a relationship between the reward and E-LAM's policy.
To do so, we let $\pi_\theta(\hat{\mathbf{s}}|\mathbf{p})$ represent the E-LAM policy with parameters $\theta$, which computes the probability of generating a predicted subjective vector $\hat{\mathbf{s}}$ given the prompt $\mathbf{p}$. We also define a reference policy $\pi_\text{ref}(\hat{\mathbf{s}}|\mathbf{p})$, typically initialized from a pre-trained E-LAM. The key insight is that an optimal policy should assign higher probability to more appropriate subjective vectors, which can be formalized as:
\begin{equation}
    \pi_\theta(\hat{\mathbf{s}}|\mathbf{p}) \propto \pi_\text{ref}(\hat{\mathbf{s}}|\mathbf{p}) \cdot \exp(r(\mathbf{p}, \hat{\mathbf{s}})/\beta),
    \label{Eq28}
\end{equation}
where $\beta$ is a temperature parameter controlling how strongly the policy favors higher-reward outputs. Rearranging (\ref{Eq28}), we can express the reward function in terms of the policies:
\begin{equation}
    r(\mathbf{p}, \hat{\mathbf{s}}) = \beta \log \frac{\pi_\theta(\hat{\mathbf{s}}|\mathbf{p})}{\pi_\text{ref}(\hat{\mathbf{s}}|\mathbf{p})} + \beta \log Z(\mathbf{p}),
\end{equation}
where $Z(\mathbf{p})$ means a normalization constant that depends only on the prompt $\mathbf{p}$. When comparing $\mathbf{s}_p$ and $\mathbf{s}_c$ for the same prompt, this constant term cancels out, i.e.,
\begin{equation}
    r(\mathbf{p}, \mathbf{s}_p) \!-\! r(\mathbf{p}, \mathbf{s}_c) = \beta \log \frac{\pi_\theta(\mathbf{s}_p|\mathbf{p})}{\pi_\text{ref}(\mathbf{s}_p|\mathbf{p})} - \beta \log \frac{\pi_\theta(\mathbf{s}_c|\mathbf{p})}{\pi_\text{ref}(\mathbf{s}_c|\mathbf{p})}.
\end{equation}
Using this implicit reward difference, we define the IoKD loss function as the negative log-likelihood of the observed preferences, i.e.,
\begin{equation}
    \mathcal{L}_\text{IoKD} = -\mathbb{E}_{\mathcal{D}} \left[ \log \sigma \left( \beta \log \frac{\pi_\theta(\mathbf{s}_p|\mathbf{p})}{\pi_\text{ref}(\mathbf{s}_p|\mathbf{p})} - \beta \log \frac{\pi_\theta(\mathbf{s}_c|\mathbf{p})}{\pi_\text{ref}(\mathbf{s}_c|\mathbf{p})} \right) \right],
\end{equation}
where $\pi_\theta$ represents the E-LAM policy being optimized. 

Then, we develop a weighted approach to adjust the temperature value $\beta$ based on the distribution characteristics of subjective vectors.
Before the first distillation round, we calculate the average subjective vector of the entire dataset, i.e., $\bar{\mathbf{s}} = \frac{1}{|\mathcal{D}_\text{IoKD}|} \sum_{i=1}^{|\mathcal{D}_\text{IoKD}|} \mathbf{s}_i$.
For each training round, the cosine similarity between the current subjective vector $\mathbf{s}_p$ and $\bar{\mathbf{s}}$ is calculated by (\ref{Eq24}).
Then, $\beta$ is dynamically set as
\begin{equation}
\beta = \beta_{\text{base}} \cdot \left(1 - \lambda \,(\text{Dis}(\mathbf{s}_p, \bar{\mathbf{s}}))\right),
\end{equation}
where $\beta_{\text{base}}$ is the default value and $\lambda$ is a scaling factor.
This adjustment mechanism is particularly suitable for numerical vector predictions, as it helps the knowledge distillation process focus on application-specific preference patterns while mitigating the influence of outliers. By assigning a larger $\beta$ to samples that closely align with the application's average preference pattern, we amplify the learning signal for these representative cases. Hence, E-LAM can effectively capture the central tendency of user preferences in the target application domain. In contrast, potential outliers that deviate significantly from the average receive a lower $\beta$, reducing their impact on the overall learning process.
\begin{figure*}[tbp!]
  \centering
  \includegraphics[width=0.95\textwidth]{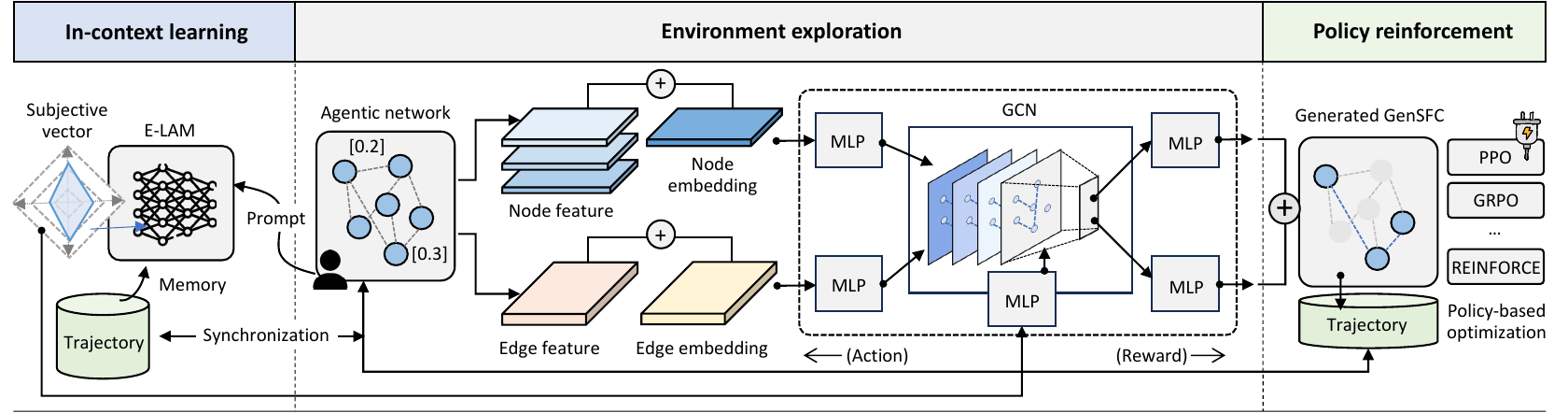} 
  \vspace{-0.12cm}
  \caption{The illustration of SRL architecture. Note that the policy optimization module is pluggable and can be implemented by diverse policy-based RL algorithms.} 
  \vspace{-0.5cm}
  \label{Figure4}
\end{figure*}


With $\mathcal{L}_{\text{IoKD}}$ and $\beta$ being configured, we elaborate on the IoKD process. First, we initialize the E-LAM with a smaller architecture than C-LAM, such as Falcon-1B and LLaMA-7B. The reference model $\pi_\text{ref}$ is initialized with the same parameters as the E-LAM to establish a stable starting point.
During distillation, IoKD samples batches repeatedly from the constructed dataset $\mathcal{D}_\text{IoKD}$. 
For each sample, we compute $\beta$ and $\mathcal{L}_\text{IoKD}$, thus updating the E-LAM parameters. 
This preference-based learning approach guides the E-LAM in distinguishing appropriate from inappropriate intent mappings while remaining computationally efficient.

\section{Symbiotic Reinforcement Learning for Agentic Network Optimization}\label{SectionV}
In this section, we present SRL to solve the problem formulated in (\ref{eq:opt_a}).
First, we introduce SRL's architecture and components.
Then, the process of SRL to achieve intent-aware network optimization is discussed, including environment exploration and in-context learning.

\subsection{Learning Architecture Overview}
Fig. \ref{Figure4} illustrates the SRL architecture, which includes three modules, i.e., environment exploration, policy optimization, and intent understanding.
Specifically, the environment exploration module leverages a Graph Convolutional Network (GCN) \cite{10606483} to effectively encode the graphical network states. 
The policy optimization module maintains a policy network that determines the action according to the specific state representation. 
This module interacts with the agentic network by performing actions and receiving rewards, which indicate the action's desirability. 
Using gradient-based optimization techniques \cite{10032267}, the policy network progressively learns to map states to high-quality GenSFCs and E-LAM selections that maximize the expected accumulated reward.

Moreover, E-LAM contributes to perceiving the environment. As shown in Fig. \ref{Figure4}, it analyzes the user intents expressed in natural language and generates the corresponding subjective vector $\mathbf{s}$. 
Similar to network state features, $\mathbf{s}$ is also injected into the environment exploration module, forming a comprehensive state representation. 
Therefore, policy optimization is guided not only by QoE factors but also by subjective intents. Consequently, the generated policies become inherently intent-aware, capable of adapting GenSFC compositions and E-LAM selections to serve specific users.
Finally, E-LAM maintains a contextual memory to reinforce itself during training. 
The overall learning architecture thus exhibits a symbiotic relationship that deeply integrates the intent understanding capabilities of LAMs and numerical optimization of DRL.

\subsection{MDP Design}
We formulate the joint optimization problem defined in (\ref{eq:opt_a}) as a Markov Decision Process (MDP). Specifically, the key MDP components are defined as follows.

\subsubsection{State Space}
Since the agentic network is modeled as an undirected graph $\mathcal{G} = (\mathcal{V}, \mathcal{E})$, the state features are associated with both nodes and edges.
Moreover, the state space also incorporates the user's subjective preference vector $\mathbf{s}$ predicted by the E-LAM. 
Consequently, the complete state space is formally defined as:
\vspace{-0.1cm}
\begin{subequations}
    \begin{flalign}
        \mathcal{S} &\triangleq (\mathcal{S}_V, \mathcal{S}_E, \mathbf{s}), \\
        \mathcal{S}_V &\triangleq \left(s^1_V = \{C_1,\,L_1,\,|O_1|,\,|F_1|\}, s^2_V, \dots, s^N_V\right),\\
        \mathcal{S}_E &\triangleq \left(s^{\{1, 2\}}_E = \{\mathbb{E}[\textit{BER}_2]\}, s^{\{1, 3\}}_E, \dots, s^{\{N, N-1 \}}_E\right),
    \end{flalign}
\end{subequations}
where $\mathcal{S}_V$ and $\mathcal{S}_E$ denote node and edge features, respectively.

\subsubsection{Action Space} The action space \( \mathcal{A} \) consists of two components: all candidate nodes (and associated edges) to compose the target GenSFC, and all available E-LAMs for user intent understanding. 
Note that to generate an \( n \)-step GenSFC, each episode is divided into \( n \) successive decision steps. At each step, SRL extends the current GenSFC by selecting a new node. Meanwhile, the selected E-LAM translates the user intent into subjective preference vectors that guide the next step. Therefore, each action \( a \in \mathcal{A} \) can be represented as \( (V_{\text{GenSFC}}, \,E_{\text{GenSFC}},\, \kappa) \), where \( V_{\text{GenSFC}} \in \boldsymbol{V} \) denotes the selected node, \( E_{\text{GenSFC}} \) is the edge formed by linking it to the predecessor, and \( \kappa \) indicates the selected E-LAM.

\subsubsection{Reward Function}
To effectively guide the direction of policy optimization, we design a hierarchical reward function comprising an immediate reward \( r_{\text{step}} \) and a final performance bonus \( r_{\text{episode}} \). At each step, SRL receives \( r_{\text{step}} \), which evaluates the desirability of adding a specific node to the GenSFC. Given that agent \( A_i \) is selected, \( r_{\text{step}} \) is defined as:
\begin{equation}
    r_{\text{step}} = r_{\text{flag}} + \mathbf{s} \cdot \left[ \mathcal{I}(C_{i}), \mathcal{I}(\textit{BER}_{i}), \mathcal{I}(L_{i}), \mathcal{I}(\frac{|F_i|}{|O_i|}) \right],
\end{equation}
where \( r_{\text{flag}} = \delta \) if \( A_i \) satisfies the functional requirement at this step, and \( -\delta \) otherwise. \( C_{i} \), \( L_{i} \), and \( \frac{|F_i|}{|O_i|} \) represent the capability, service latency, and crash probability of \( A_i \), respectively. \( \textit{BER}_{i} \) denotes the BER of the communication link between \( A_i \) and its predecessor; for the first node, \( \textit{BER}_{i} = 0 \). \( \mathcal{I}(\cdot) \) denotes a normalization function that mitigates the influence of raw value magnitudes to ensure stable learning. By weighting these intent-relevant factors using the predicted user preference vector \( \mathbf{s} \), SRL is incentivized to select nodes that better align with subjective service expectations.
Upon completion of the full GenSFC, the environment evaluates its overall quality and issues a performance-based bonus \( r_{\text{episode}} \).
Based on (\ref{eq:opt_a}), $r_\text{episode}$ is defined as (\ref{eq35}).
\begin{figure*}
\begin{equation}
r_{\text{episode}} = (1 - \omega_P P_{\text{outage}}) \left( (1 - \omega_B B_{\text{GenSFC}}) \ln\left( \frac{\omega_C C_{\text{GenSFC}}}{C_{\text{th}}} \right) - \ln(\omega_L L_{\text{GenSFC}}) \right) - \mathcal{F}(\kappa).
\label{eq35}
\end{equation}
\hrule
\end{figure*}
This hierarchical reward structure provides both step-wise guidance and global alignment with user intent, enabling SRL to progressively make informed decisions to maximize user QoE.

\subsection{Environment Exploration and Policy Optimization}
We implement the exploration and learning process using a GCN-enhanced PPO framework. In this part, we detail how environmental states are encoded, how actions are selected, and how the policy is optimized.

\subsubsection{Graph-Based State Encoding} Recall that at each step $t$, the state $s_t$ is formatted as $(\mathcal{S}_V,\, \mathcal{S}_E,\, \mathbf{s})$. Let $\mathbf{X} \in \mathbb{R}^{N \times |\mathcal{S}_V|}$ denote the node feature matrix, and $\mathbf{A} \in \mathbb{R}^{N \times N \times |\mathcal{S}_E|}$ denote the binary adjacency matrix. We adopt a multi-layer GCN to embed the spatially-correlated topology. Each GCN layer performs neighborhood aggregation as:
\begin{equation}
    \mathbf{H}^{(l+1)} = \sigma\left( \hat{\mathbf{D}}^{-1/2} \hat{\mathbf{A}} \hat{\mathbf{D}}^{-1/2} \mathbf{H}^{(l)} \mathbf{W}^{(l)} \right),
\end{equation}
where $\hat{\mathbf{A}} = \mathbf{A} + \mathbf{I}$, $\hat{\mathbf{D}}$ is the degree matrix of $\hat{\mathbf{A}}$, $\mathbf{W}^{(l)}$ is a learnable weight matrix, and $\sigma(\cdot)$ is the ReLU activation. The initial layer input is $\mathbf{H}^{(0)} = \mathbf{X}$, and multiple such layers propagate contextual information across graph neighborhoods.

After the final GCN layer, a global mean pooling is applied over node embeddings, i.e.,
\begin{equation}
\mathbf{h}_{\text{graph}} = \frac{1}{N} \sum_{i=1}^N \mathbf{H}^{(L)}_i.
\end{equation}
$\mathbf{h}_{\text{graph}}$ is concatenated with two auxiliary embeddings: a user intent embedding $\mathbf{h}_{\text{intent}} = \phi(\hat{\mathbf{s}})$ and an E-LAM index embedding $\mathbf{h}_{\text{model}} = \psi(m_t)$. These components form the latent representation $\mathbf{z}_t = [\mathbf{h}_{\text{graph}} \,\|\, \mathbf{h}_{\text{intent}} \,\|\, \mathbf{h}_{\text{model}}]$, where $\phi(\cdot)$ and $\psi(\cdot)$ are fully connected layers with non-linear activations.

\subsubsection{Policy Inference and Action Selection} The policy is represented by a stochastic mapping $\pi_\varphi(a_t | s_t)$, parameterized by $\varphi$, that assigns a probability distribution over actions given the current state. The latent representation $\mathbf{z}_t$ is processed by the actor network to produce this distribution over the discrete action space $\mathcal{A}$, i.e.,
\begin{equation}
    \pi_\varphi(a_t|s_t) = \text{Categorical}(\text{MLP}_{\text{actor}}(\mathbf{z}_t)),
\end{equation}
where MLP denotes a Multilayer Perceptron module \cite{10606483} for encoding. The critic network, which shares the encoder, estimates the state value function:
\begin{equation}
V_{\varphi'}(s_t) = \text{MLP}_{\text{critic}}(\mathbf{z}_t).
\end{equation}
In step $t$, an action $a_t$ is sampled from $\pi_\varphi$, and the environment returns the reward $r_\text{step}$ and the next state $s_{t+1}$.

\subsubsection{Policy Gradient and Optimization} 
To optimize policy, our objective is to maximize the expected cumulative reward:
\begin{equation}
\mathcal{J}(\varphi) = \mathbb{E}_{\pi_\varphi} \left[ \sum_{t=0}^{T} \gamma^t r_t \right].
\end{equation}
Using the policy gradient theorem, the gradient of this objective can be written as:
\begin{equation}
\nabla_\varphi J(\varphi) = \mathbb{E}_{\pi_\varphi} \left[ \nabla_\varphi \log \pi_\varphi(a_t|s_t) \cdot \hat{A}_t \right],
\end{equation}
where $\hat{A}_t$ denotes the advantage estimate indicating how much better an action is compared to the expected baseline. According to \cite{10032267}, $\hat{A}_t$ is defined as
\begin{equation}
\hat{A}_t = \sum_{l=0}^{T-t} (\gamma \lambda)^l \delta_{t+l}, \, \delta_t = r_t + \gamma V_{\varphi'}(s_{t+1}) - V_{\varphi'}(s_t).
\end{equation}

Finally, we utilize PPO to optimize the policy within a trust region. The clipped surrogate objective is defined as
\begin{equation}
\mathcal{L}^{\text{PPO}}(\varphi) = \mathbb{E}_t \left[ \min\left( r_t(\varphi) \hat{A}_t, \text{clip}(r_t(\varphi), 1 - \epsilon, 1 + \epsilon) \hat{A}_t \right) \right],
\end{equation}
where $r_t(\varphi) = \frac{\pi_\varphi(a_t|s_t)}{\pi_{\varphi{\text{old}}}(a_t|s_t)}$ is the importance sampling ratio, and $\epsilon$ is a hyperparameter that controls the clipping range.
The policy can then be optimized by gradient ascent using $\mathcal{L}^{\text{PPO}}(\varphi)$.
Note that LAMeTA regards this part as a pluggable module.
Other policy-based RL algorithms, such as REINFORCE and PPO variants, can also be applied.

\subsection{In-context Learning of E-LAM}
While E-LAM provides RL with intent understanding capabilities, RL also enhances E-LAM through environmental feedback. 
Specifically, E-LAM maintains a structured context memory $\mathcal{M}$ containing historical records of prompt, predicted subjective vector, and resulting reward, i.e.,
\begin{equation}
\mathcal{M} = \{m_i = \langle \mathbf{p}_i, \mathbf{s}_i, r_i \rangle \}_{i=1}^{M}.
\end{equation}
At each step $t$ with prompt $\mathbf{p}_t$, E-LAM retrieves the $k$ most recent examples from $\mathcal{M}$ and analyzes the consistency of preference vectors and reward trends.
If fetched subjective vectors maintain high similarity and rewards exhibit a positive trend, this indicates that SRL is operating within a consistent intent domain and producing increasingly effective solutions. In such cases, E-LAM leverages this pattern consistency to reinforce its prediction confidence, i.e.,
\begin{equation}
s_t = \text{E-LAM}(p_t | \{m_{t-1}, m_{t-2}, \ldots, m_{t-k}\})
\end{equation}
Conversely, when E-LAM detects a significant divergence in the generated subjective vector compared to recent examples, it implements an automatic calibration mechanism. This divergence may indicate either an outlier user prompt or a potential misinterpretation. To maintain robust performance, E-LAM employs a weighted averaging approach that moderates potentially outlier translations:
\begin{equation}
s_t^{\text{calibrated}} = \iota \cdot s_t + (1-\iota
) \cdot \bar{s},
\end{equation}
where $\bar{s}$ represents the average of recent subjective vectors in $\mathcal{M}$, and $\iota \in [0,1]$ is a calibration factor that decreases as divergence increases. This calibration ensures that subjective vectors remain within reasonable bounds of the application-specific preference distribution even when facing outliers.
Additionally, E-LAM implements a recovery mechanism for cases where it fails to generate valid preference vectors. In such instances, it falls back to $\bar{s}$ for one round.

Through in-context learning, E-LAM achieves adaptive intent understanding without requiring expensive retraining, effectively utilizing the reinforcement learning outcomes to enhance its own capabilities.

\begin{figure*}[tbp!]
  \centering
  \includegraphics[width=0.8\textwidth]{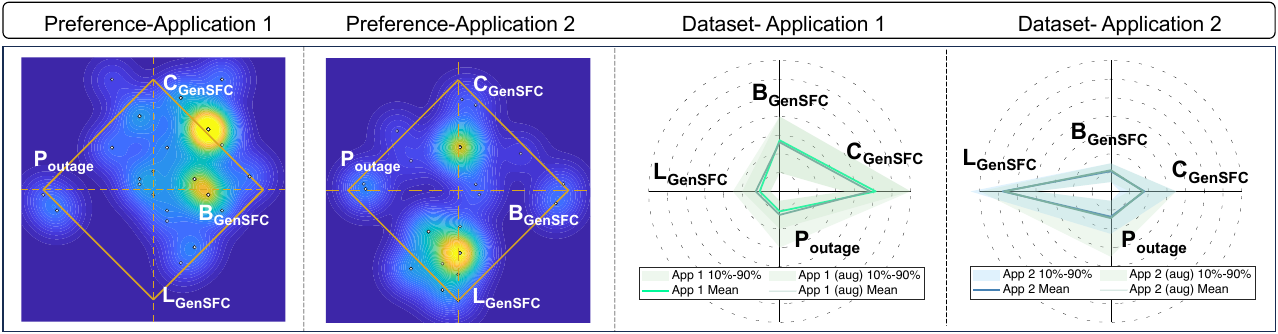} 
  \caption{The illustration of user preferences for two applications and the effectiveness of dataset augmentation.} 
  \vspace{-0.2cm}
  \label{Figure5}
\end{figure*}

\section{Numerical Results}\label{SectionVI}
In this section, we implement the proposed LAMeTA. 
Then, we perform extensive experiments to answer 1) whether IoKD can precisely translate user natural language intent into subjective preferences and 2) whether SRL can efficiently compose GenSFCs and select E-LAMs for users according to their specific applications and maximize the QoE.

\textbf{Testbed}: The experiments are conducted on a server with three NVIDIA RTX A5000 GPUs (each has 24 GB of memory) and an AMD Ryzen Threadripper PRO 3975WX 32-core CPU with 263 GB of RAM in Ubuntu 20.04 LTS.

\textbf{Experimental Settings}: We implement LAMeTA and simulate an agentic network consisting of 81 agents, each associated with one of four service types, denoted as type-1 through type-4. In our setup, service requests involve the sequential composition of three service types, i.e., type-1, type-2, and type-3. We apply GPT-4.5\footnote{Model available at: https://openai.com/index/gpt-4/} as C-LAM. For E-LAMs, we apply two variants from Deepseek: a 7-billion-parameter (7B) model and a 1.5-billion-parameter (1.5B) model\footnote{Model available at: https://huggingface.co/deepseek-ai}.

\vspace{-0.2cm}
\subsection{Preference Visualization}
To investigate how user preferences vary in different applications, we construct a benchmark dataset of 100 human-annotated $\langle \textit{prompt}, \textit{subjective factor} \rangle$ pairs. The prompts that contain user intent are sampled from two distinct applications: one involving multimedia content generation and another one targeting question-answering. We embed the resulting subjective vectors and visualize the aggregated user preferences as heatmaps in Fig.~\ref{Figure5}. For Application 1, users predominantly prioritize $C_\text{GenSFC}$ and also emphasize $B_\text{GenSFC}$, indicating a strong emphasis on the received content quality. In contrast, Application 2 focuses on $L_\text{GenSFC}$ and $C_\text{GenSFC}$, reflecting the demand to rapidly obtain high-quality answers.

For each application, we randomly select 10 samples as few-shot demonstrations and retain the remaining 40 samples as test sets. The C-LAM is then prompted to synthesize $\langle \textit{prompt},$ $ \textit{subjective factor} \rangle$ pairs, resulting in two augmented datasets with 500 samples. Fig.~\ref{Figure5}(right) compares the distribution of real and synthesized preference vectors. We can observe that synthetic samples exhibit high semantic similarity to real data, confirming the reliability of C-LAM in preserving task-specific subjective semantics during data augmentation.
\begin{figure*}[tpb]
\centering
\begin{minipage}{\textwidth}
	\centering
	\begin{minipage}{0.42\textwidth}
	\includegraphics[width=\textwidth]{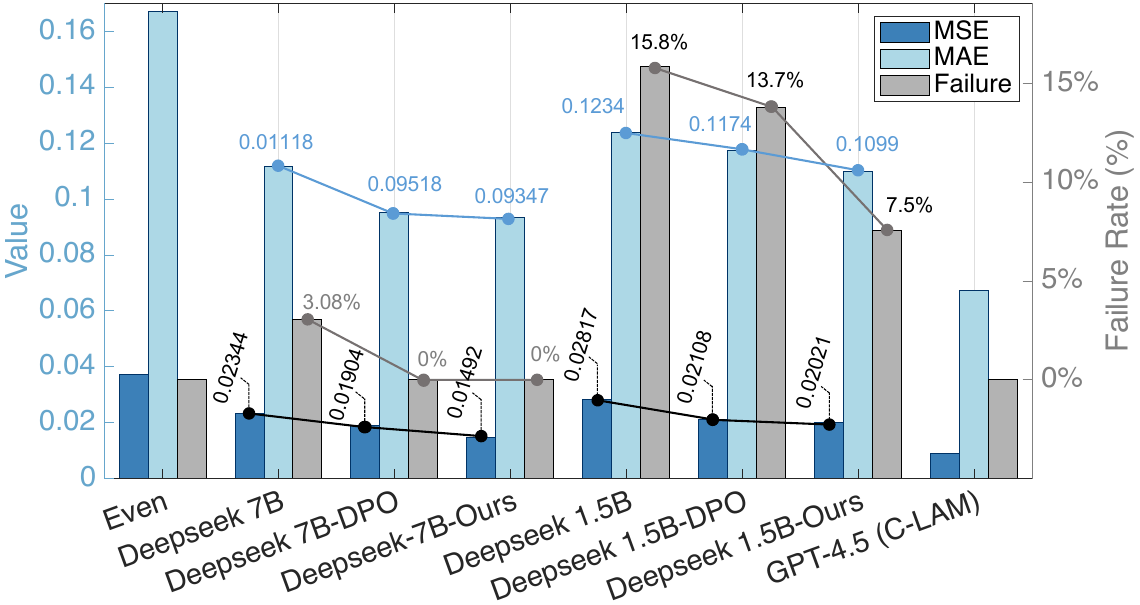}
	\subcaption*{(a) The precision of Application 1.}
	\end{minipage}
	\begin{minipage}{0.42\textwidth}
	\includegraphics[width=\textwidth]{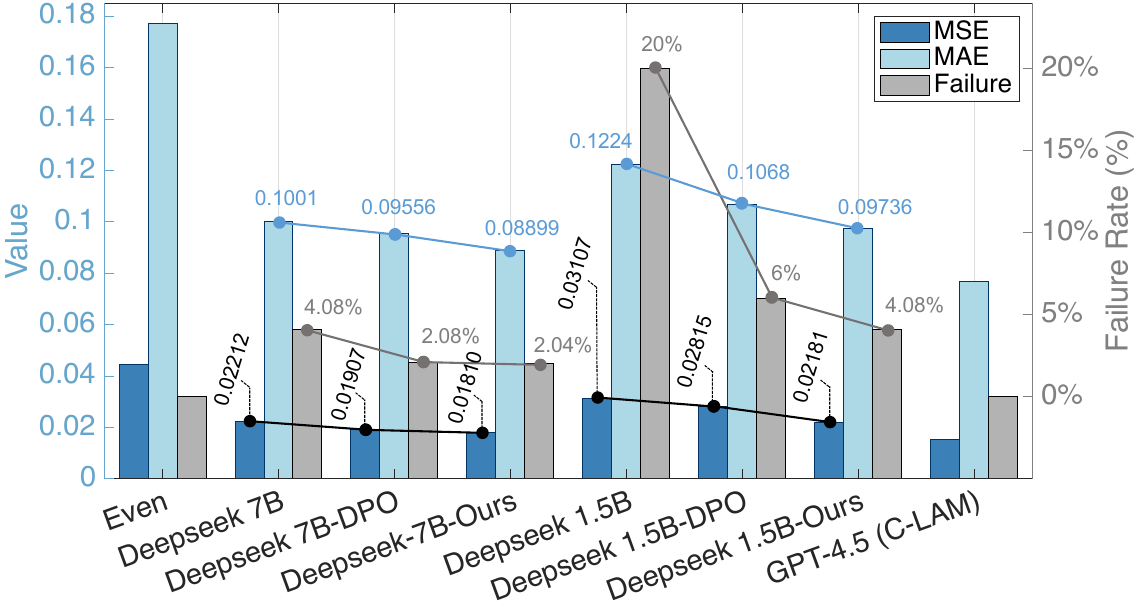}
    \subcaption*{(b) The precision of Application 2.}
	\end{minipage}
        \caption{The precision of user intent translation. Note that the precision is evaluated by averaging the prediction results over 40 test samples for each application. We can observe that IoKD facilitates E-LAMs to approach the performance of C-LAM.}
	\label{Figure6}
\end{minipage}
 \vspace{-0.3cm}
\end{figure*}
\begin{figure}[htbp!]
  \centering
  \includegraphics[width=0.45\textwidth]{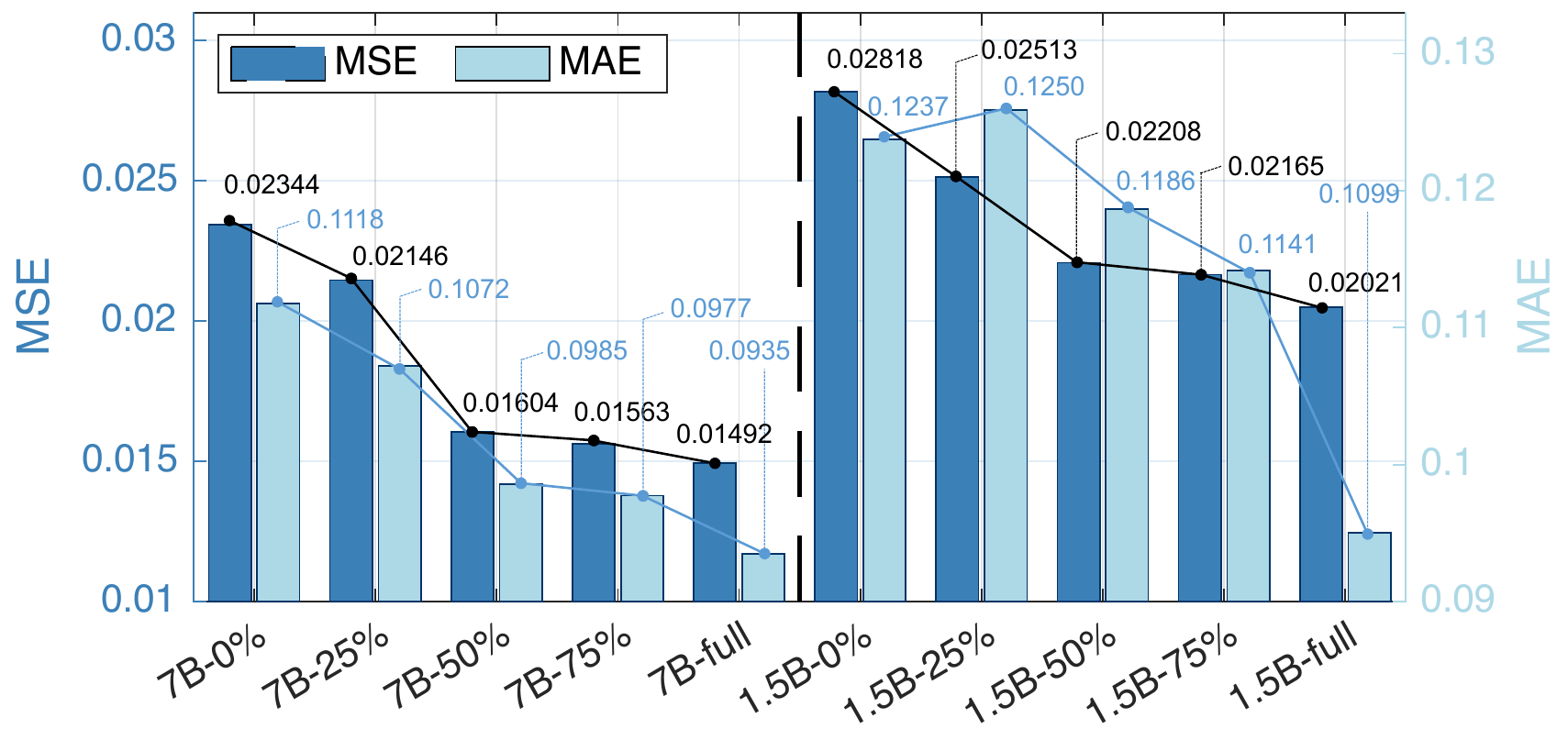}
  \caption{The impact of dataset sizes for IoKD on intent understanding precision.}
  \vspace{-0.5cm}
  \label{Figure7}
\end{figure}

\vspace{-0.2cm}
\subsection{Precision of User Intent Understanding}
We perform IoKD and evaluate the precision of E-LAMs for intent understanding.
Specifically, two metrics are used to measure the prediction error, namely Mean Absolute Error (MAE) and Mean Squared Error (MSE).
In addition, we report the failure rate, defined as the proportion that LAM fails to generate a qualified subjective vector $\mathbf{s}$.
As shown in Fig. \ref{Figure6}, we begin with a default baseline that applies an even preference vector $\mathbf{s} = [0.25, 0.25, 0.25, 0.25]$ regardless of intent. 
We can observe that this setting leads to the highest MAE and MSE across both applications, as it fails to capture any meaningful user-specific variation.
Next, we evaluate raw E-LAMs without knowledge distillation. 
Fig. \ref{Figure6} shows that the 7B variant achieves an MSE of 0.0234 and 0.0282 in Applications~1 and 2, respectively, while the 1.5B variant shows slightly higher error due to its limited capacity.
Then, we apply DPO to distill the intent understanding capability from the C-LAM to each E-LAM. 
The distilled models exhibit substantial improvements: MSE drops to 0.190 for the 7B model and 0.211 for the 1.5B model in Application~1.
In addition, failure rates are reduced by more than 50\% in most cases.
Finally, the proposed IoKD achieves the best performance. 
By integrating weighted loss, we further reduce MSE by 21.6\% (for 7B) and 4.1\% (for 1.5B) in Application~1, and 5.1\% (for 7B) and 22.5\% (for 1.5B) in Application~2, compared with the original DPO. 
This improvement demonstrates the efficiency of IoKD in guiding E-LAM to learn intent predictions.

Furthermore, we evaluate the precision of the IoKD with a varying number of samples.
As shown in Fig. \ref{Figure7}, increasing the number of samples for IoKD keeps increasing the resulting precision, which further demonstrates the effectiveness of our data augmentation algorithm.
\begin{figure*}[tpb]
\centering
\begin{minipage}{\textwidth}
	\centering
	\begin{minipage}{0.4\textwidth}
	\includegraphics[width=\textwidth]{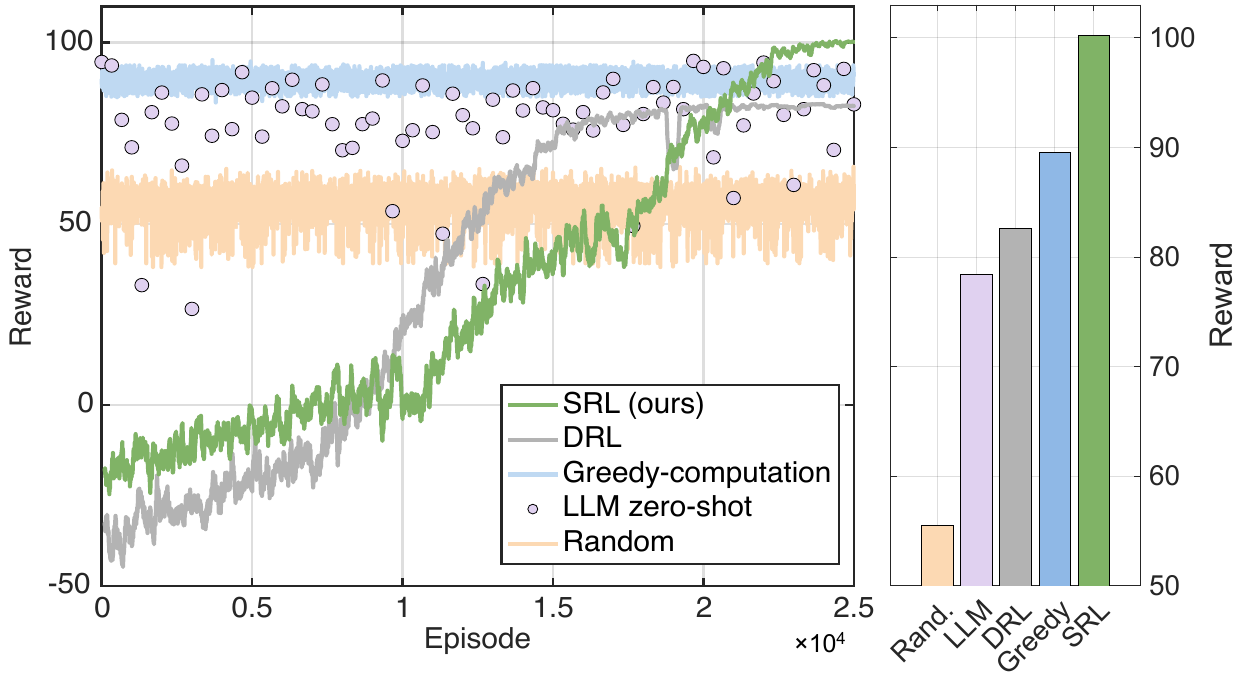}
	\subcaption*{(a) The performance of training on Application 1.}
	\end{minipage}
	\begin{minipage}{0.39\textwidth}
	\includegraphics[width=\textwidth]{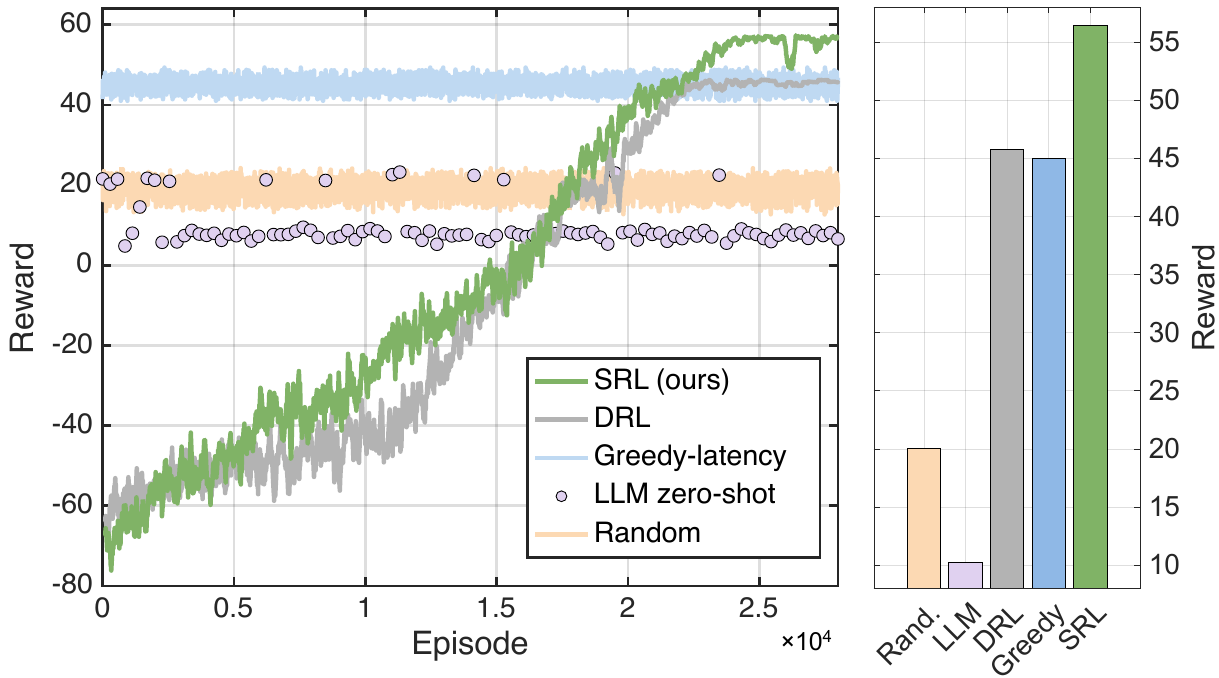}
        \subcaption*{(b) The performance of training on Application 2.}
	\end{minipage}
        \caption{The convergence and performance of SRL and baselines for agentic network optimization.}
	\label{Figure8}
\end{minipage}
 \vspace{-0.2cm}
\end{figure*}
\begin{figure*}[htbp]
  \centering
  \includegraphics[width=0.75\textwidth]{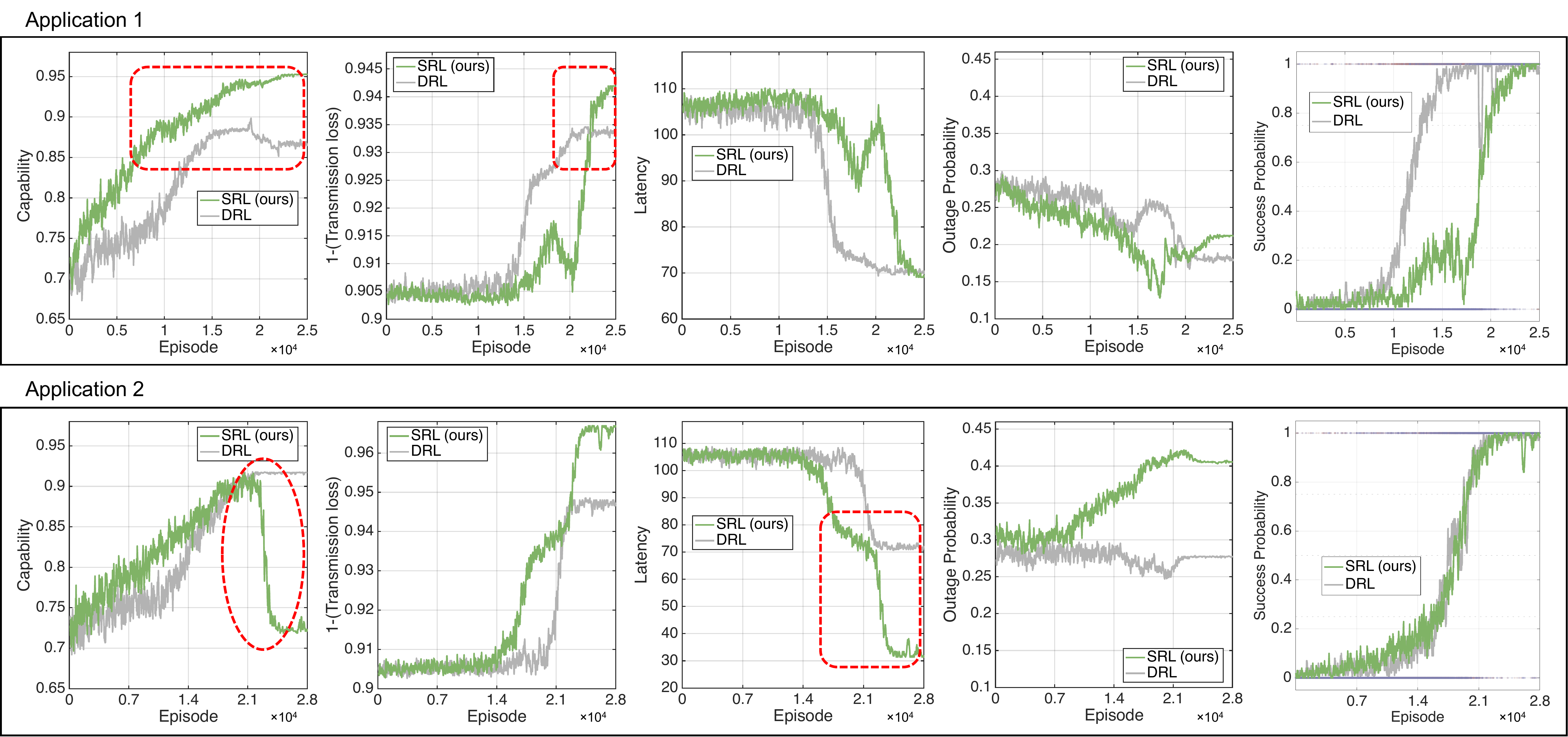}
  \caption{The trends of each QoE factor during training for Applications 1 (up) and 2 (bottom), respectively.} 
  \vspace{-0.3cm}
  \label{Figure9}
\end{figure*}
\subsection{Efficiency of GenSFC Compositions}
\subsubsection{Convergence and Acquired Reward}
We evaluate SRL performance for GenSFC composition and E-LAM selection across the aforementioned applications. 
As shown in Fig.~\ref{Figure8}, we compare five baselines: i) \textbf{Random}, which composes GenSFCs by randomly selecting agents; ii) \textbf{Greedy}, which sequentially selects nodes that maximize the most dominant subjective factor in $\mathbf{s}$; iii) \textbf{DRL (Even)}, a PPO-based baseline that assumes even user preferences; iv) \textbf{LLM-zero shot}, which queries GPT-4 with the full system model, user intent, and task prompt to generate GenSFCs end-to-end; and v) \textbf{SRL}, which integrates intent-aware decision-making.

As shown in Fig.~\ref{Figure8}(a), Random suffers from severe instability and low utility due to the expansive action space and heterogeneous agent configurations. Similarly, generic DRL, which lacks user-specific guidance, demonstrates inferior performance, as it cannot prioritize the factors that users concern about. Although Greedy performs better by aggressively optimizing for the most dominant preference dimension, its unidimensional strategy overlooks trade-offs across other factors, ultimately yielding suboptimal GenSFCs. 
The LLM zero-shot generation shows inconsistent results. In Application~1, it partially understands the user preference for capability and achieves comparable performance to Greedy. However, in Application~2, it fails to minimize the service latency, leading to significant QoE degradation.
In contrast, the proposed SRL consistently outperforms all baselines. It achieves performance improvements of 17.2\% and 23.5\% over the best-performing alternatives in Applications 1 and 2, respectively. 
This improvement demonstrates the effectiveness of applying E-LAMs to provide explicit learning signals that guide the policy to align with subjective QoE objectives.

\subsubsection{Performance on Each QoE Factor}
To further interpret the intent-aware behavior of SRL, we track the trend of each QoE factor during training. As shown in Fig.~\ref{Figure9}, in Application~1, where capability and BER are highly weighted, SRL steadily increases $C_{\text{GenSFC}}$ while simultaneously reducing $B_{\text{GenSFC}}$. This trend indicates its ability to identify both high-performing nodes and low-noise communication paths, ensuring generation quality and semantic fidelity. Although the outage probability $P_{\text{outage}}$ remains moderately high, it is less important due to its small weight in $\mathbf{s}$, which justifies its sacrifice in favor of more critical service aspects.
In Application~2, users prioritize low latency. SRL rapidly reduces $L_{\text{GenSFC}}$ by favoring lightweight agents with shorter service times and minimal transmission delays. This latency-driven optimization comes at the cost of a reduction in $C_{\text{GenSFC}}$, since lightweight nodes typically have lower generation capability. Nevertheless, since $\omega_L > \omega_C$, the reduction in capability does not impede subjective QoE. The overall reward remains consistently higher than that of all baselines, confirming that SRL effectively balances competing objectives according to personalized intent vectors.
 
\begin{figure}[tbp!]
  \centering
  \includegraphics[width=0.25\textwidth]{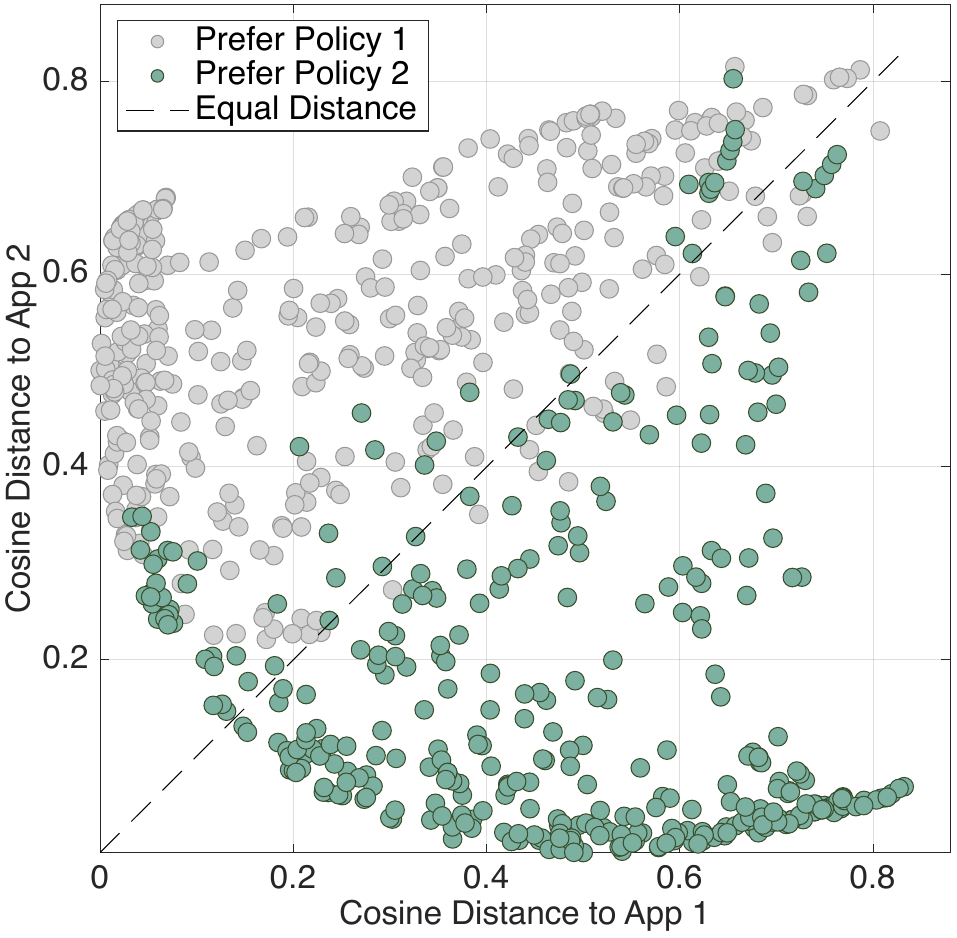}
  \vspace{-0.1cm}
  \caption{The illustration of preferred policy for users with different intents. Note that the distance is measured by cosine similarity and is defined in (\ref{Eq24}).} 
  \label{Figure10}
  \vspace{-0.65cm}
\end{figure}

\subsubsection{Intent-Aware GenSFC Composition}
Here, we evaluate the capacity of SRL to customize GenSFCs for users with diverse intents. 
We simulate 800 users, each associated with a subjective preference vector. 
For each user, we compute the cosine similarity between its preference vector and the average preference vectors of the two applications. 
We then apply both trained policies and assess which one achieves a higher reward.
As depicted in Fig.~\ref{Figure10}, over 90\% of users can maximize their QoE by selecting the policy trained on the application closest to their personal intent.
This alignment confirms that SRL-generated policies encapsulate application-specific patterns.
Therefore, LAMeTA allows users to simply select the SRL policy trained by the most similar samples, thus maximizing subjective QoE. This architecture generalizes the agentic network from rigid, fixed-function optimization to flexible, user-driven service provisioning.

\begin{figure}[tbp!]
  \centering
  \includegraphics[width=0.5\textwidth]{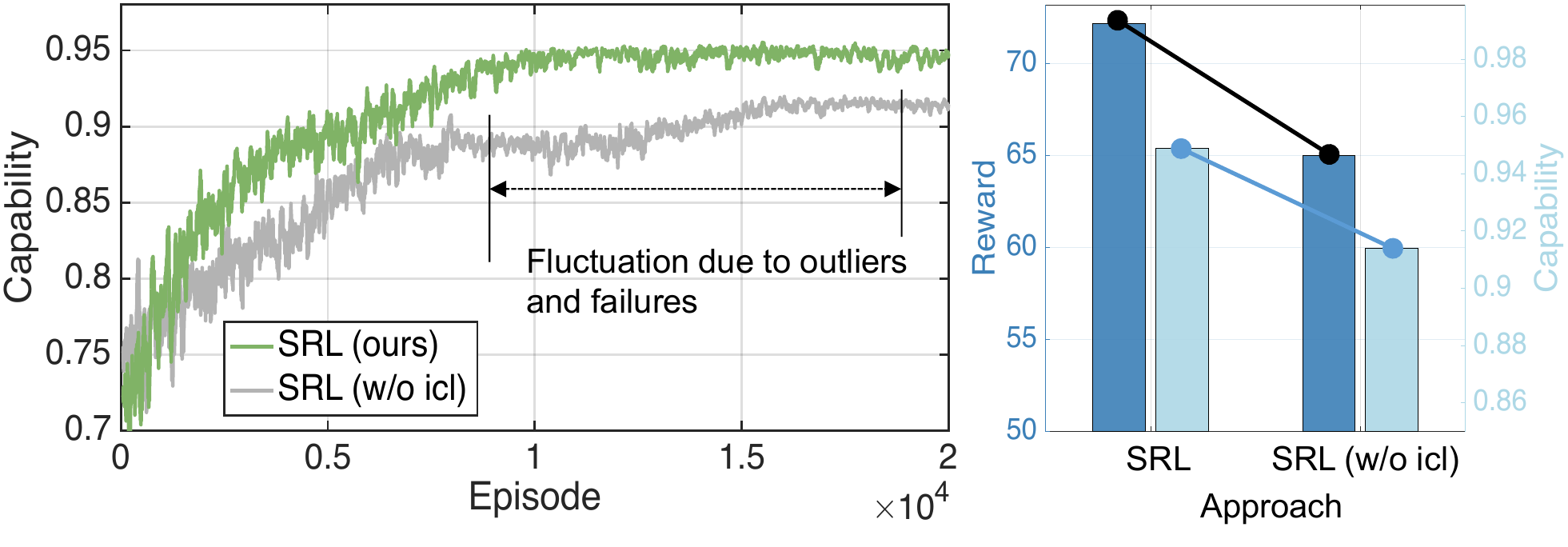}
  \vspace{-0.6cm}
  \caption{The ablation study to evaluate the effectiveness of in-context learning.} 
  \label{Figure11}
  \vspace{-0.5cm}
\end{figure}
\subsubsection{Inspection on In-context Learning}
Fig.~\ref{Figure11} demonstrates the validity of in-context learning in SRL. Specifically, we train a policy targeting Application 1 while inserting 20\% samples from Application 2 to simulate outliers and failures. We can observe that SRL consistently achieves higher capability scores than the variant without in-context learning, with a final capability of 0.95 compared to 0.91. Moreover, SRL converges more rapidly, reaching the 0.90 threshold approximately 40\% faster. This result reveals that outlier detection of in-context learning contributes significantly to preventing error propagation and stabilizing training. 

\section{Conclusion}\label{SectionVII}
This paper has proposed LAMeTA for intent-aware agentic network optimization, which contains two stages named IoKD and SRL.
Specifically, IoKD distills intent understanding capability from C-LAMs to user-friendly and customized E-LAMs. 
SRL achieves the seamless integration of E-LAM and DRL to efficiently solve the joint optimization problem on GenSFC composition and E-LAM selection.
Experimental results have validated the effectiveness of the proposed approach in maximizing QoE with diverse user intents.

\bibliographystyle{IEEEtran}
\bibliography{Ref}

\end{document}